\documentclass[aps,amsmath,amssymb, nofootinbib,  twocolumn, 
superscriptaddress,prx]{revtex4-1}

\usepackage{graphicx}
\usepackage{color}
\usepackage{bbold}

\newcommand{\zb}{{\boldsymbol z}}

\newcommand{\bit}{\begin{itemize}}
\newcommand{\eit}{\end{itemize}}

\newcommand{\f}{\frac}
\renewcommand{\>}{\right\rangle}
\newcommand{\<}{\left\langle}
\newcommand{\ba}{\begin{align}}
\newcommand{\ea}{\end{align}}
\newcommand{\be}{\begin{equation}}
\newcommand{\ee}{\end{equation}}
\newcommand{\bi}{\begin{itemize}}
\newcommand{\ei}{\end{itemize}}
\newcommand{\lf}{\left(}
\newcommand{\ri}{\right)}
\newcommand{\dd}{\mathrm{d}}

\newcommand{\Tr}{\operatorname{Tr}}

\newcommand{\lag}{\mathcal{L}}
\newcommand{\rp}{\mathrm{RP}}
\newcommand{\cp}{\mathrm{CP}}
\newcommand{\nccp}{\mathrm{NCCP}}

\newcommand{\CC}{\mathcal{C}}

\newcommand{\spn}{\mathcal{N}_s}
\newcommand{\etav}{\eta_\text{VBS}}
\newcommand{\etan}{\eta_\text{N\'eel}}

\def\CC#1{\left[#1\right]}

\def\de{\text{d}}

\def\+{\dagger}

\def\avrg#1{\left\langle#1\right\rangle}

\def\s{\sigma}

\graphicspath{{DCPfigs/}}

\begin{document}

\newcommand{\bra}[1]{\< #1 \right|}
\newcommand{\ket}[1]{\left| #1 \>}

\title{Deconfined Quantum Criticality, Scaling Violations,  and Classical Loop Models}
\author{Adam Nahum} 
\affiliation{Department of Physics, Massachusetts Institute of Technology, Cambridge, MA 02139, USA}
\author{J. T. Chalker}
\affiliation{Theoretical Physics, Oxford University, 1 Keble Road, Oxford OX1 3NP, United Kingdom}
\author{P. Serna}
\affiliation{Theoretical Physics, Oxford University, 1 Keble Road, Oxford OX1 3NP, United Kingdom}
\affiliation{Departamento de F\'isica -- CIOyN, Universidad de Murcia, Murcia 30.071, Spain}
\author{M. Ortu\~no}
\author{A. M. Somoza}
\affiliation{Departamento de F\'isica -- CIOyN, Universidad de Murcia, Murcia 30.071, Spain}
\date{\today}

\begin{abstract}
\noindent
Numerical studies of the transition between N\'eel and valence bond solid phases in 2D quantum antiferromagnets give strong evidence for the remarkable scenario of deconfined criticality, but display strong violations of finite-size scaling that are not yet understood. We show how to realise the universal physics of the N\'eel--VBS transition in a 3D classical loop model (this model  includes the subtle interference effect that suppresses hedgehog defects in the N\'eel order parameter).  We use the loop model  for simulations of  unprecedentedly large systems (up to linear size $L=512$). Our results are compatible with a continuous transition at which both N\'eel and VBS order parameters are critical, and we do not see conventional signs of first order behaviour. However, we show that the scaling violations are stronger than previously realised and are \emph{incompatible} with conventional finite-size scaling over the range of sizes studied, even if allowance is made for a weakly or marginally irrelevant scaling variable.  In particular, different approaches to determining the anomalous dimensions  $\eta_\text{VBS}$ and $\eta_\text{N\'eel}$ yield very different results. The assumption of conventional finite-size scaling leads to estimates which drift to negative values at large sizes, in violation of the unitarity bounds. In contrast, the decay with distance of critical correlators on scales much smaller than system size is consistent with large positive anomalous dimensions. Barring an unexpected reversal in behaviour at still larger sizes, this implies that the transition, if continuous, must show unconventional finite-size scaling, for example from an additional dangerously irrelevant scaling variable. Another possibility is an anomalously weak first order transition. By analysing the renormalisation group flows for the non-compact $\cp^{n-1}$ field theory (the $n$-component Abelian Higgs model) between two and four dimensions, we give the simplest scenario by which an anomalously weak first order transition can arise without fine-tuning of the Hamiltonian.
\end{abstract}


\maketitle

\section{Introduction}

\noindent
The paradigmatic `deconfined' quantum phase transition is that separating the N\'eel antiferromagnet from the columnar valence bond solid (VBS) for a square lattice of spin--1/2s. The theoretical arguments of Refs.~\cite{deconfined critical points, quantum criticality beyond, critically defined} indicate that the N\'eel-VBS phase transition is described by the noncompact $\cp^1$ ($\nccp^1$) model \cite{Motrunich Vishwanath}, a field theory with bosonic spinons $\zb=(z_1, z_2)$ coupled to a noncompact $\mathrm{U}(1)$ gauge field:
\be\label{NCCP1 lagrangian}
\mathcal{L} = |(\nabla - i A) \zb|^2 + \kappa (\nabla \times A)^2 + \mu |\zb|^2 + \lambda |\zb|^4.
\ee
This theory is defined in three-dimensional Euclidean spacetime; the N\'eel order parameter  is proportional to $\zb^\dag \vec \sigma \zb$, where $\vec \sigma$ are the Pauli matrices.

Numerical results for the J--Q model (the Heisenberg antiferromagnet supplemented with a four-spin interaction \cite{Sandvik JQ}) support the validity of this continuum description \cite{Sandvik logs, Banerjee et al, lou sandvik kawashima, Kawashima deconfined criticality, melko kaul fan, pujari damle alet honeycomb}, as does work on the $\mathrm{SU}(n)$ generalisation of the problem at large $n$ \cite{kaul sandvik large n, Monopole scaling dim, Kaul Sachdev large n, Monopole scaling dim 2}.   Unfortunately though, the existence of a continuous phase transition in both the $\nccp^1$ model and the $\mathrm{SU}(2)$ lattice magnets remains a vexed question. While simulations of the J-Q model are compatible with a direct continuous transition, they show strong violations of finite size scaling \cite{Sandvik logs, Kawashima deconfined criticality, Jiang et al, deconfined criticality flow JQ}. These persist up to the largest system sizes studied so far, and hamper the extraction of meaningful critical exponents \cite{Kawashima deconfined criticality}. Additionally, direct numerical studies of the lattice $\nccp^1$ field theory have disagreed as to whether the transition is continuous \cite{Motrunich Vishwanath, Motrunich Vishwanath 2}, or whether scaling violations similar to those seen in the lattice magnets should be interpreted as the initial stages of runaway flow to a first order transition \cite{Kuklov et al}. 

Are the scaling violations seen at the N\'eel-VBS transition indeed signs of a first order transition, with an anomalously large correlation length \cite{Kuklov et al, Jiang et al, deconfined criticality flow JQ}, or are they due to the critical theory possessing a weakly irrelevant scaling variable \cite{Sandvik logs, Bartosch, Banerjee et al SU(3)}, or do they indicate something more exotic? This issue remains controversial. Its relevance extends beyond quantum magnets, since the critical behaviour of the $\nccp^1$ model is important for various other fundamental problems in statistical mechanics. For example, this field theory is believed to describe the three-dimensional classical $\mathrm{O}(3)$ model when hedgehog defects are disallowed \cite{Motrunich Vishwanath}, as well as the columnar ordering transition in the classical dimer model on the cubic lattice \cite{PowellChalkerPRL, Charrier Alet Pujol, PowellChalkerPRB, Chen et al, Alet extended dimer}. (In the latter example $\mathrm{SU}(2)$ symmetry is absent microscopically, but argued to emerge at the critical point.) There is also numerical evidence that similar scaling violations afflict the $\mathrm{SU}(3)$ and $\mathrm{SU}(4)$ generalisations of the deconfined transition \cite{Kaul SU(3) SU(4), Kawashima deconfined criticality}.

In this paper we introduce a new model which is ideally suited for studying the universal features of the N\'eel-VBS transition, and perform simulations on very large systems (of linear size up to $512$ lattice spacings, and $640$ for a few selected observables). We verify that the model shows the basic features expected from the $\nccp^1$ field theory (Eq.~\ref{NCCP1 lagrangian}): an apparently continuous direct transition, with emergent $\mathrm{U}(1)$ symmetry for rotations of the VBS order parameter at the critical point. However, we show that scaling violations are even stronger than previously appreciated. Conventional finite-size scaling assumptions are not obeyed: the data cannot be made to show scaling collapse, and quantities that would normally be expected to be universal instead drift with system size. The larger sizes considered here show that these drifts are stronger than the logarithmic form conjectured previously \cite{Sandvik logs, Banerjee et al}.

In common with Ref.~\cite{Kawashima deconfined criticality}, we see a drift in finite-size estimates of critical exponents.  We show that this is more drastic than  previously apparent.  Estimates of the anomalous dimensions of both the N\'eel and VBS order parameters, as extracted from the correlation functions $G(r)$ at distances $r$ comparable with the system size (e.g. $r=L/2$) yield negative values at large sizes. Negative anomalous dimensions are ruled out for a conformally invariant critical point by the unitarity bounds \cite{unitarity bounds 1, unitarity bounds 2}. On the other hand, the decay of $G(r)$ with $r$ for $r\ll L$ appears consistent with the large positive anomalous dimensions suggested for a deconfined critical point.  It is conceivable that the transition could be continuous, but that conventional finite-size scaling could fail as a result of a dangerously irrelevant variable \cite{Kaul SU(3) SU(4)}. For example, in this scenario correlators $G(r)$ with $1\ll r\ll L$ would presumably show the true positive anomalous dimensions, while correlators with $r$ of order $L$ would behave anomalously (as in e.g. $\phi^4$ theory above 4D \cite{brezin zinn justin fss}). The hypothetical dangerously irrelevant variable discussed here should not be confused with the much discussed $\mathbb{Z}_4$ anisotropy for the VBS order parameter (Sec.~\ref{emergent symmetry section}) which is dangerously irrelevant in a different sense.

Therefore --- unless there is a reversal of the drift in exponents at still larger sizes, which seems unlikely --- there are two  possibilities: either the transition is continuous with unconventional finite-size scaling behaviour (for example as a result of a dangerously irrelevant scaling variable) or it is first order. We will discuss both possibilities but cannot rule out either. We do not see the conventional signs of a first order transition, such as double-peaked probability distributions for the energy and other quantities.

On the other hand, an alternative hypothesis put forward previously --- that the scaling violations are due simply to a weakly or marginally irrelevant scaling variable \cite{Sandvik logs, Banerjee et al, Bartosch} --- is \emph{not} supported by  our data. We also rule out any explanation in terms of unconventional dynamic scaling, i.e. deviations from dynamical exponent $z=1$: our model has $z=1$ by construction since it is isotropic in three dimensions. This isotropy is also a convenient feature from the point of view of simulations.

Turning to theory, we analyse the topology of the RG flows in the $\nccp^{n-1}$ model between two and four dimensions, in order to assess the possibility of an anomalously weak first order transition. This analysis unifies what is known about this field theory in $4-\epsilon$ dimensions, in $2+\epsilon$ dimensions, and at large $n$, and extends previous partial results \cite{March Russell}. Treating $n$ as continuously varying, we argue that in 3D there is a universal value $n_*$ below which the deconfined critical point disappears by merging with a tricritical point. The argument does not fix the value of  $n_*$, which could be greater or smaller than two, so does not tell us whether the $\nccp^1$ model has a continuous transition. However, it does have the following consequences. \emph{If} $n_*$ happens to be greater than two,  there is a possible mechanism by which a very weak first order transition can appear for a range of $n$ values (i.e. a large correlation length can be obtained without the need for fine tuning of the Hamiltonian). The argument also shows that $n_*$ is greater than one. Therefore the inverted XY transition in the model with $n=1$ is \emph{not} analytically connected to the critical point in the large-$n$ regime of the $\nccp^{n-1}$ model, contrary to assumptions made in previous work.

This picture for the RG flows also clarifies that the usual $2+\epsilon$ expansion of the $\mathrm{O}(3)$ sigma model does not describe the conventional $\mathrm{O}(3)$ transition in 3D, but rather the deconfined critical point (if it exists at $n=2$).  This is natural:  hedgehogs are crucial in determining the critical behaviour of the $\mathrm{O}(3)$ model in 3D  \cite{Motrunich Vishwanath}, and the $2+\epsilon$ expansion presumably fails to account for them. The conclusion is also in line with the RG result that the $2+\epsilon$ approach to the 3D $\mathrm{O}(M)$ model should fail  when $M$ is less than a universal value $M_c$, conjectured to be above 3, as a result of neglecting the topology of the sphere \cite{Cardy Hamber}. Interestingly, our best estimates for the correlation length exponent at the deconfined transition (Sec.~\ref{corr exp drift}) are close to $\nu = 1/2$, smaller than most previous estimates but in good agreement with the $2+\epsilon$ predictions (Sec.~\ref{failure of 2+epsilon}).

Returning to lattice models, our numerical strategy is,  instead of focussing on a simple two-dimensional quantum Hamiltonian, to construct a simple 3D classical model that is well adapted to large-scale Monte Carlo simulations. While the correspondence with classical lattice models in one dimension higher is a standard tool for studying quantum phase transitions, one might at first glance think that this tool is not available for deconfined criticality. This is because deconfinement relies crucially on the fact that the Euclidean action for the spins in 2+1 dimensions --- \emph{unlike} the energy functional for a classical spin model in 3 dimensions --- contains imaginary terms (Berry phases). The effect of these terms is to endow hedgehogs in the N\'eel order parameter with position-dependent complex fugacities \cite{Haldane 2+1, Read Sachdev VBS and spin peierls, Read Sachdev VBS and spin peierls 2}. After coarse-graining, this leads to a phase cancellation effect that effectively suppresses hedgehogs \cite{deconfined critical points, quantum criticality beyond, critically defined}.

Contrary to the naive expectation above, we show that the remarkable physics of deconfinement, including the suppression of hedgehogs by phase cancellation, {is} present in our 3D classical model. This model is formulated in terms of configurations of loops on a lattice and is a variant of the models of Refs.~\cite{cpn loops short, cpn loops long}. The loop configurations have positive Boltzmann weights, so define a conventional classical statistical mechanics problem. However, the partition function can also be mapped to a lattice field theory for $\cp^{n-1}$ spins, and in this representation the Boltzmann weights are \emph{not} necessarily real. We show by a direct calculation that they include the complex hedgehog fugacities necessary for deconfined criticality.

The loop model introduced here has qualitative features in common with loop ensembles arising in worldline quantum Monte Carlo techniques for sign-problem free Hamiltonians such as the J-Q model \cite{Kaul et al QMC review}. However, direct simulation of a quantum Hamiltonian leads to an ensemble of worldlines in continuous imaginary time, whereas the loop model is an isotropic three-dimensional lattice model. This is a desirable feature for numerical simulations as it fixes an otherwise unknown velocity and eliminates a potentially significant source of corrections to scaling \cite{anisotropy footnote}. The geometric form of the model also motivates new observables  --- for example, we find it useful to consider some percolation--like observables such as the number of system-spanning strands and the fractal dimension of the loops.

\section{Loop Model}
\label{loop model section}

An astonishing variety of critical phenomena can be studied using classical loop gases. The present lattice model involves two species (colours) of loops, or $n$ colours in the $\mathrm{SU}(n)$ generalisation.  It has a phase in which infinite loops proliferate, and one in which all loops are short. The short-loop phase spontaneously breaks lattice symmetry because the system must choose between four symmetry-related ways to pack the short loops. 

The transfer matrix for loop models of this kind gives a correspondence with a 2D quantum magnet on the square lattice \cite{cpn loops long}. The colour of a strand is related to the state of the spin (at a given point in Euclidean space-time) in the quantum problem. The infinite-loop phase corresponds to the N\'eel phase: the presence of infinite loops is equivalent to the presence of long-range spin correlations.  The four degenerate short-loop phases map to the four equivalent columnar VBS patterns on the square lattice.  The schematic correspondence between the loop model and the continuum field theory (Eq.~\ref{NCCP1 lagrangian}) is that the two species of loops are worldlines of the two species of bosonic spinons $(z_1, z_2)$. (See Sec.~\ref{continuum description section} for more detail on the continuum limit.)

\begin{figure}
\includegraphics[width=\linewidth]{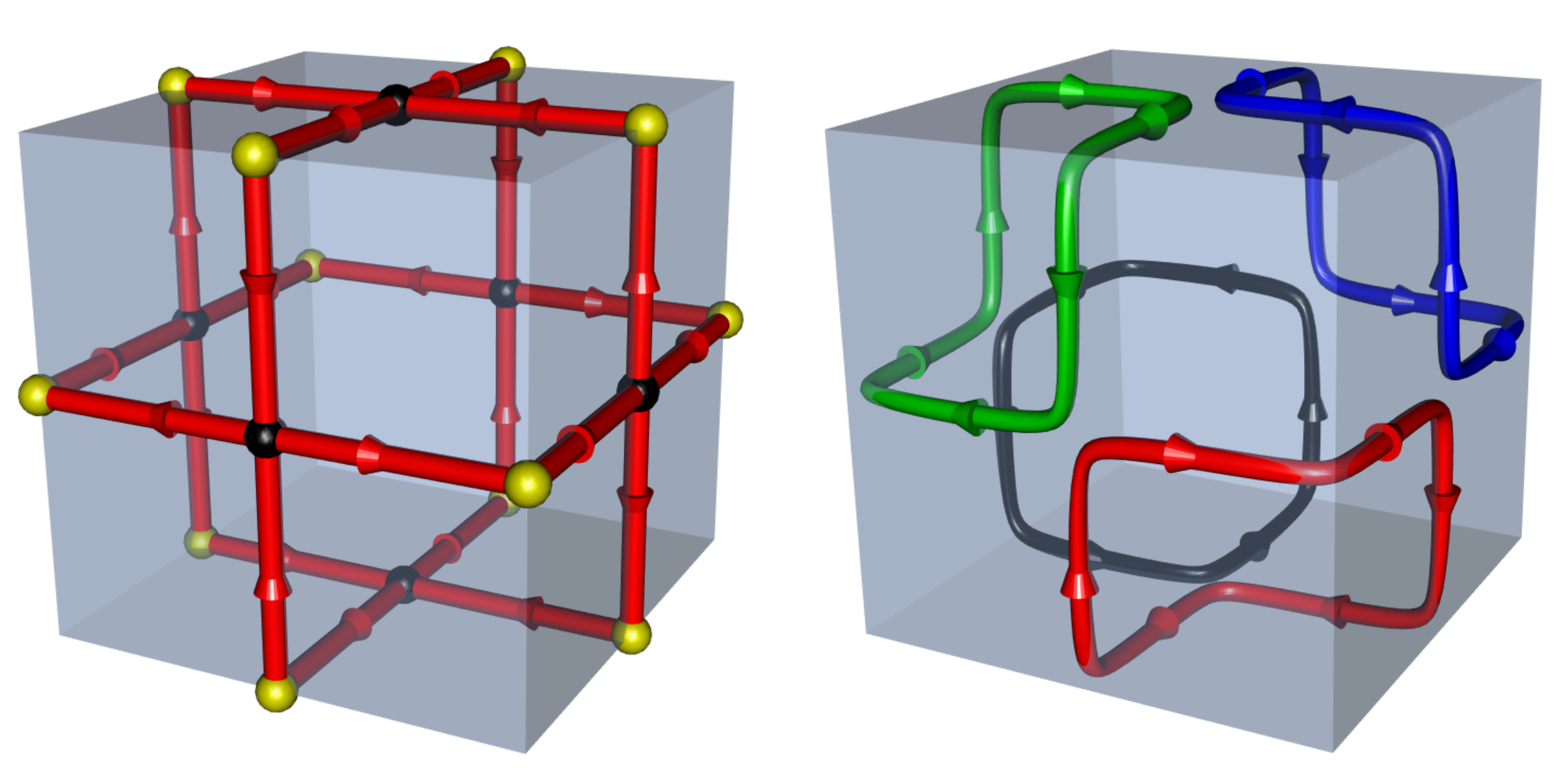}
 \caption{Structure of L lattice. Left: Nodes and links (with associated orientations) in the unit cell of the L-lattice. The two sublattices of nodes are indicated. Right: One of the four equivalent packings of minimal--length loops on this lattice.}
 \label{L lattice figure}
\end{figure}

The loop model is a modification of those studied in Refs.~\cite{cpn loops short, cpn loops long}, with an additional interaction chosen to drive the model through a transition without explicitly breaking the symmetry between the four short-loop (VBS) states.  The model lives on a four-coordinated lattice with cubic symmetry proposed by Cardy \cite{cardy network models review}. This `3D L lattice' is shown in Fig.~\ref{L lattice figure} (left). Formally it can be defined by starting with two interpenetrating cubic lattices, $C_1$ and $C_2$, with lattice spacing $2$, displaced from each other by $(1,1,1)$:
\ba\label{C1 and C2}
C_1 &= (2\mathbb{Z})^3, & C_2 & = (2\mathbb{Z}+1)^3.
\end{align}
The faces of $C_1$ intersect the faces of $C_2$ along lines: these define the links of the L lattice. The L lattice is bipartite; its two sublattices are marked in yellow and black in Fig.~\ref{L lattice figure}.

\begin{figure}[b]
\includegraphics[width=\linewidth]{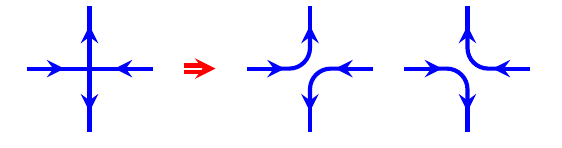}
 \caption{The two possible configurations of a node ($\sigma_r = \pm 1$).}
 \label{nodes}
\end{figure}

We orient the links of the L lattice such that each node has two incoming and two outgoing links, with the two incoming links parallel and the two outgoing links parallel, as in Figs.~\ref{L lattice figure},~\ref{nodes}. This assignment is unique up to a reversal of all orientations.

Breaking up each L lattice node by pairing the links in one of the two ways shown in Fig.~\ref{nodes} gives a completely-packed loop configuration. In the simplest case, the partition function is just the equal weight sum over all such  configurations, with one of $n$ colours assigned to each loop (the case of main interest here is $n=2$):
\be\label{preliminary partition function}
Z = \sum_{\substack{\text{coloured} \\ \text{loop configs}}} 1.
\ee
We can of course perform the sum over colours explicitly to give $Z = \sum_{\substack{\text{loop} \\ \text{configs}}} n^\text{no. loops}$. Note that the loops are automatically consistently oriented, since the pairing of links at a node is always between an incoming and an outgoing link.

With the above Boltzmann weight, the system is (for $n\leq 4$) in a phase where infinite loops proliferate \cite{cpn loops short, cpn loops long}. We wish to add an interaction that drives the system into a phase with only short loops. First, consider the extreme limit of such a phase, which is a configuration in which every loop has the minimal possible length (which is six links). There are only four such configurations and they are related by lattice symmetry. One is shown in Fig.~\ref{L lattice figure}, right.

A general loop configuration is determined by the binary choice of link pairing at each of the nodes. We denote this binary degree of freedom at node $r$ by an Ising-like variable $\sigma_r=\pm 1$. To fix the sign convention for $\sigma$, let us pick one of the four minimal-length configurations as a reference and declare that all $\sigma_r$ are equal to $+1$ in it. The four minimal-length configurations are then those in which all the $\sigma$s on the same sublattice ($A$ or $B$) have the same sign. We can define an order parameter of the schematic form $\vec \varphi = ( \sigma_A , \sigma_B )$ which distinguishes the four short-loop phases. This is the analogue of the VBS order parameter in the quantum problem.

We introduce an interaction between nearest-neighbour $\sigma$s on the \emph{same} sublattice (i.e. between nodes of like colour in Fig.~\ref{L lattice figure}):
\begin{gather}\label{full partition function}
Z  =  \sum_{\substack{\text{coloured} \\ \text{loop configs}}} 
\exp \lf - E \ri,
\\
E = - J \Bigg(
\sum_{\<r, r'\> \in A}  \hspace{-1mm} \sigma_r \sigma_{r'} + \hspace{-1mm}
\sum_{\<r, r'\> \in B}  \hspace{-1mm} \sigma_r \sigma_{r'} 
\Bigg).
\end{gather}
(The sum over \emph{un}coloured loop configurations is equivalent to a sum over the $\sigma$s.) With this choice, there is a direct transition at  $J_c$ between a phase that has extended loops and $\langle \vec \varphi \rangle = 0$, and one that has only short loops and $\langle \vec \varphi \rangle \neq 0$.

As we would expect from the quantum correspondence \cite{cpn loops long}, the continuum description of the above model is the $\nccp^{n-1}$ model. In Sec.~\ref{continuum description section} we show this directly by mapping the loop model to a lattice  $\cp^{n-1}$ model and coarse graining, paying special attention to the fate of hedgehogs.

\section{Overview of Results}
\label{numerics overview}

\noindent
We first summarise the salient results of our simulations.  

At the most basic level, they confirm that the loop model shows the central features of the deconfined N\'eel--VBS transition, and probes the same universal physics as the J--Q \cite{Sandvik JQ} and related quantum models.

\begin{figure}[t]
 \begin{center}
\includegraphics[width=\linewidth]{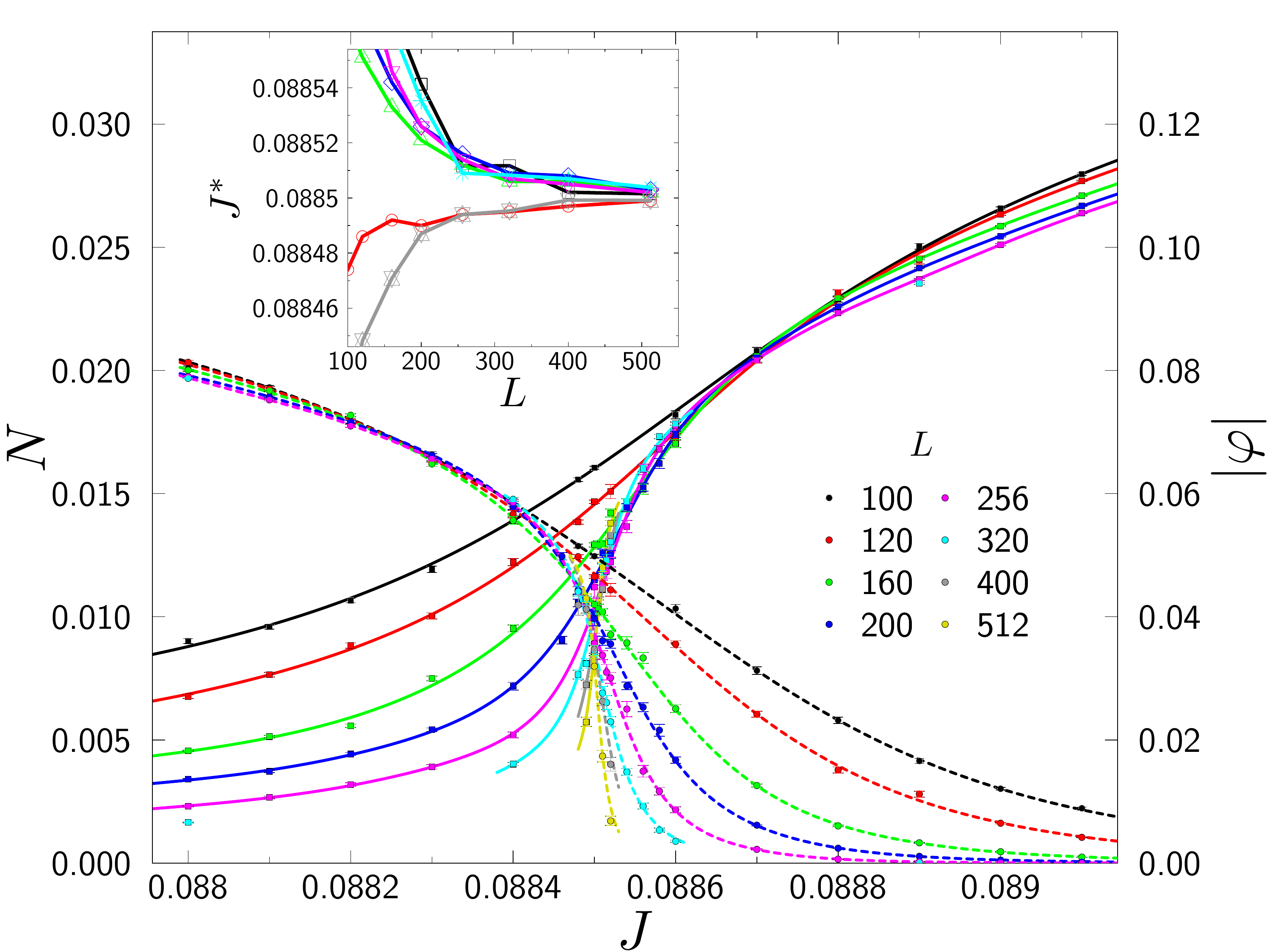}
\end{center}
\caption{N\'eel order parameter  $N$ (dashed lines) and VBS order parameter $|\vec \varphi|$ (full lines) as a function of $J$ for several system sizes $L$.  Continuous curves are interpolations using the multiple histogram method \cite{Ferrenberg}. Inset:  pseudocritical couplings $J^*(L)$ obtained from various observables [$N$ (blue rhombi), $|\vec\varphi|$ (down-pointing magenta  triangles), crossings of $4N$ \& $|\vec\varphi|$ (grey stars), $C$ (black squares), ${\cal N}_s$ (up-pointing green triangles), $B_\varphi$ (red circles) and $V$ (turquoise asterisks)].} 
\label{orderparameters}
\end{figure}

\begin{figure}[b]
 \begin{center}
 \includegraphics[width=\linewidth]{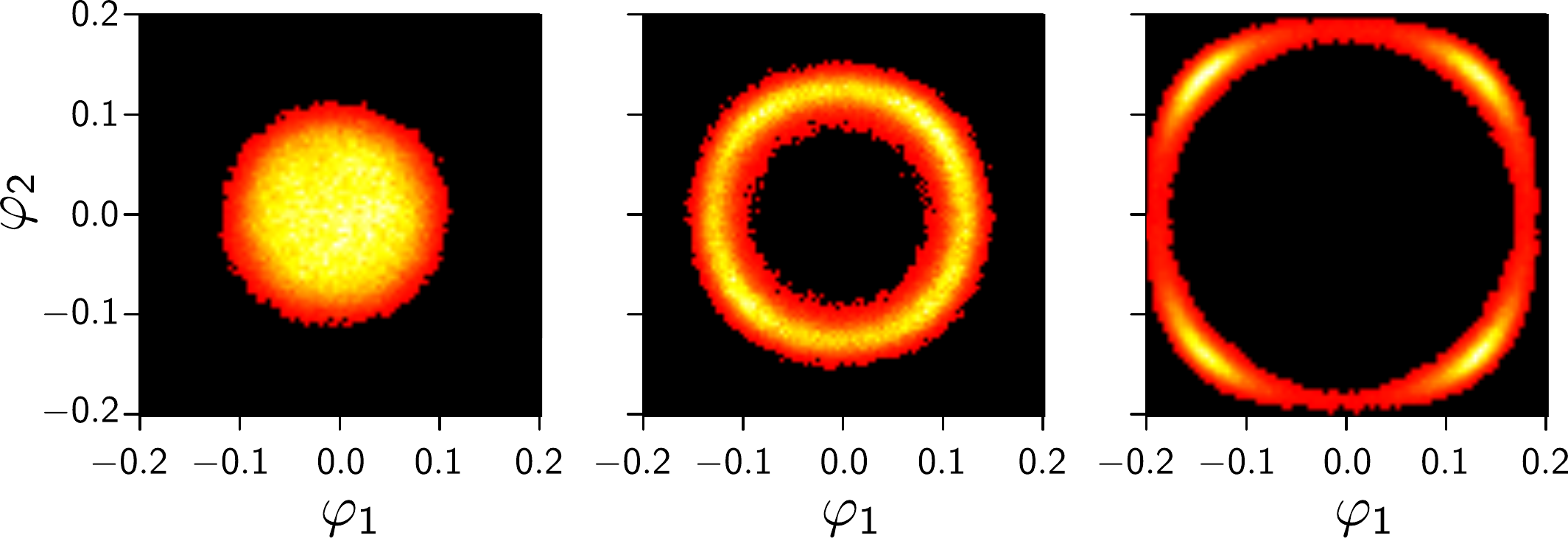}
 \end{center}
 \caption{Probability distribution of $\vec\varphi=(\varphi_1, \varphi_2)$ for size $L=64$ and (from left to right)  $J=0.0886$, $J=0.091$, and  $J=0.096$, corresponding
respectively to  the critical point, the VBS phase in the $\mathrm{U}(1)$ regime, and the VBS phase in the $\mathbb{Z}_4$ regime. The crossover from $\mathrm{U}(1)$ to $\mathbb{Z}_4$ symmetry is analysed further in Sec.~\ref{emergent symmetry section}.
 \label{fig:2ddis}}
\end{figure}

We find a direct and apparently continuous transition between N\'eel and VBS phases. Fig.~\ref{orderparameters} shows the order parameters for these phases, for various system sizes $L$, very close to the critical point (details  in Sec.~\ref{order parameters}). The data suggests a single transition. This is confirmed by examining finite--size pseudocritical couplings $J_c(L)$ determined from various observables (inset to Fig.~\ref{orderparameters});   all extrapolate to the same value as $L\rightarrow \infty$ within error bars, so we are confident there is a single transition at 
\be
J_c = 0.088501(3).
\ee
At small sizes, the  estimates of critical exponents are compatible with those found in the J--Q model at similar sizes and in direct simulations of the $\nccp^1$ model \cite{Sandvik logs,lou sandvik kawashima,Kawashima deconfined criticality, Motrunich Vishwanath 2, melko kaul fan}. As expected \cite{deconfined critical points, Sandvik JQ, lou sandvik kawashima}, we see an emergent $\mathrm{U}(1)$ symmetry for rotations of the VBS order parameter $\vec \varphi$ close to this critical point. (The emergence of this $\mathrm{U}(1)$ symmetry is equivalent to the noncompactness of the gauge field in the continuum action Eq.~\ref{NCCP1 lagrangian}  \cite{deconfined critical points,quantum criticality beyond}.) Within the VBS phase, the emergent $\mathrm{U}(1)$ symmetry survives up to a lengthscale $\xi_\text{VBS}$ that is parametrically larger than the correlation length $\xi$. As for the J--Q model \cite{lou sandvik kawashima}, the $\mathrm{U}(1)$ symmetry is apparent in the probability distribution for $\vec \varphi$; see Fig.~\ref{fig:2ddis}. 

\begin{figure}[t]
 \begin{center}
\includegraphics[width=\linewidth]{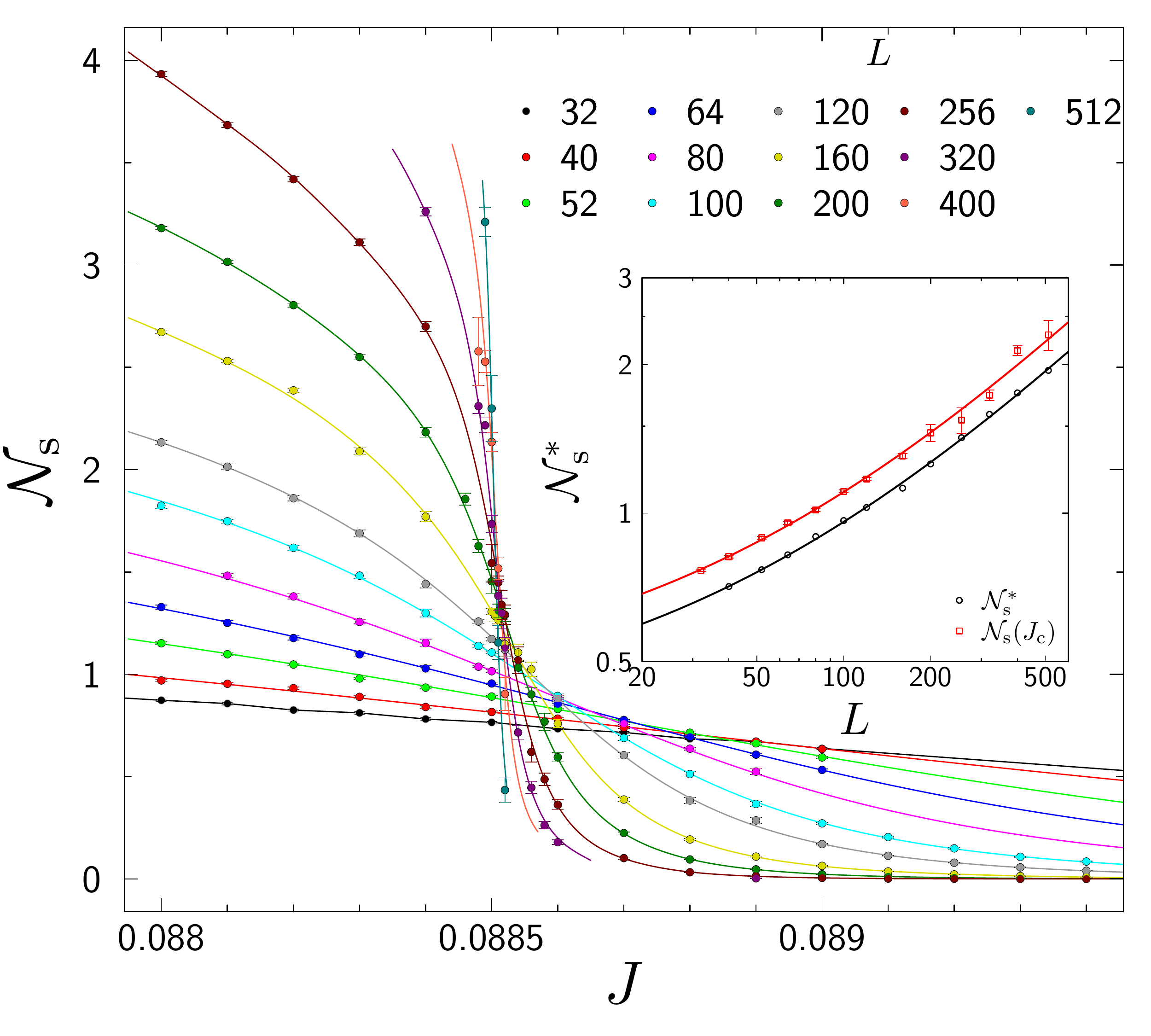} 
\end{center}
 \caption{Spanning number  $\mathcal{N}_{\rm s}$ as a function of $J$, for various $L$. Inset: $L$--dependence of critical spanning number $\mathcal{N}_\mathrm{s}^*$, defined (1) as $\mathcal{N}_{\rm s}$ at $J=J_c$ (red dots);  (2) as $\mathcal{N}_{\rm s}$ at the point where the slope $|\dd \mathcal{N}_{\rm s} / \dd J|$ is maximal   (black dots). Continuous curves are power-law fits, $\mathcal{N}_\mathrm{s}^* = A L^p+B$,  with exponents $p=0.61(5)$ and $p=0.62(3)$ respectively. Note log-log scale.\label{winding}}
\end{figure}

Despite the above features, finite--size scaling properties at the transition are anomalous in various ways. For example, an appropriately-defined stiffness for the N\'eel vector --- which would be a universal constant at a conventional critical point --- increases slowly with system size, and the critical exponents estimates also drift as the size is increased. Similar features were seen in previous numerical work on the J--Q model \cite{Sandvik JQ,Sandvik logs,Kawashima deconfined criticality, Jiang et al, deconfined criticality flow JQ}, but the larger sizes considered here show the scaling violations are stronger than previously apparent. For a detailed picture of the transition, we analyse a variety of observables. Violations of finite-size scaling are visible in almost all quantities and do not decrease as $L$ is increased. For this reason, we are unable to fit the size dependence of the data near the critical point assuming {either} scaling corrections coming from an irrelevant scaling variable (even if it is very weakly irrelevant) or logarithmic corrections similar to those considered in Refs.~\cite{Sandvik logs, Banerjee et al}. (See Secs.~\ref{order parameters},~\ref{spanning number section},~\ref{corr exp drift}.)   

Fig.~\ref{winding} shows the `spanning number' $\spn$ versus the coupling $J$ for various system sizes. $\spn$ is the average number of strands which span the system in a given direction, and is a measure of the stiffness of the N\'eel order parameter. Instead of tending to a universal value as dictated by standard finite-size scaling, the crossing points $\spn^*$ drift upwards as a power of $L$: Fig.~\ref{winding}, inset. (See Sec.~\ref{spanning number section} for details.)

We calculate the correlation length exponent $\nu$ and the anomalous dimensions $\eta_\text{N\'eel}$ and $\eta_\text{VBS}$ using several observables. In the text, results will be presented with statistical errors in the last significant digit shown in brackets in the usual way; for reasons that will be apparent, we are not generally able to estimate systematic errors.

Estimates of $\nu$ obtained from finite-size scaling analyses of different quantities are in reasonable agreement, but  drift significantly, from $\nu \gtrsim 0.6$ at small sizes to values around $\nu \sim 0.46$ for the largest sizes. In contrast, values of $\nu$ obtained from the variation of the correlation length with distance from the critical point lie in the range $0.45$ --- $0.5$ with less dependence on size (Sec.~\ref{corr exp drift}).

Strikingly, the behaviour of the correlation functions at the critical point suggests different values of the anomalous dimensions  $\eta_\text{N\'eel}$ and $\eta_\text{VBS}$ depending on the range of $r$ used to extract them. Values obtained from correlation functions at separation $L/2$ both drift from values above $0.2$ to values \emph{below} zero at large sizes. Negative $\eta$s \emph{violate} the unitarity bound $\eta\geq0$ \cite{unitarity bounds 1, unitarity bounds 2}. In contrast, there is evidence that behaviour for $r\ll L$ is consistent with positive values for the anomalous dimensions. We note that the use of correlators at separation $L/2$ to determine $\eta$ assumes finite-size scaling, which is a stronger assumption than that the continuity of the transition, as we will discuss in Sec.~\ref{correlation functions section}.

\begin{figure}
 \begin{center}
\includegraphics[width=\linewidth]{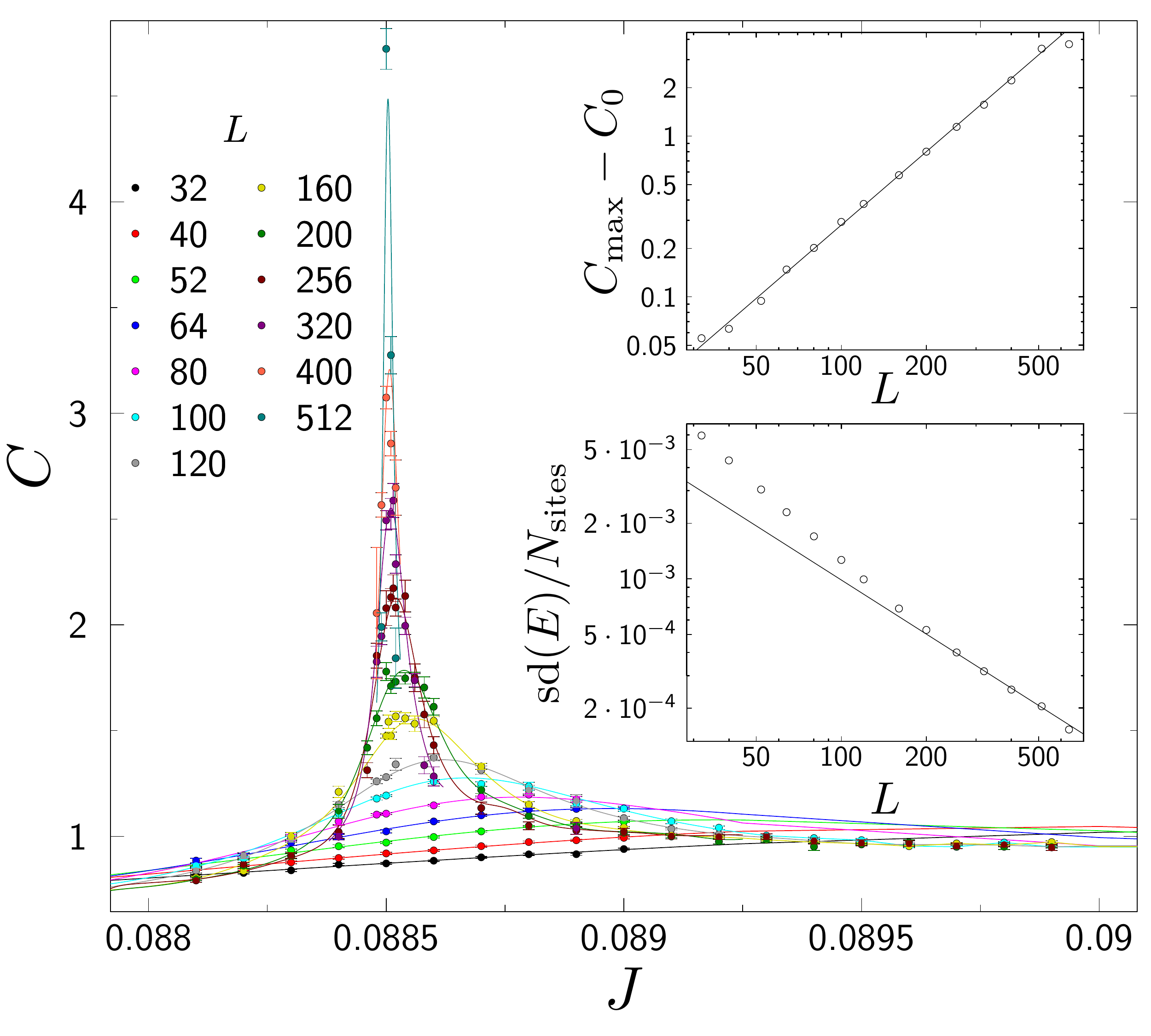}
  \end{center}
\caption{
Main Panel: heat capacity $C$ versus $J$. Upper inset:  peak height $C_\text{max}$ with constant $C_0=0.9836$ subtracted to account for background. Fit is power law with exponent $\alpha/\nu=1.52(2)$. Lower inset: standard deviation in energy at $J=0.08850$. Fit is with exponent $(\alpha/\nu-3)/2=-0.97(4)$.
\label{heat capacity plot}}
\end{figure}

In view of the above, the transition can only be continuous if some subtlety invalidates the usual finite--size--scaling expectations. (Of course in principle  there could be a drastic change in behaviour at  still larger sizes $L\gg 512$, but  the data gives no reason to expect this.) Therefore it is natural to ask whether  the transition is first order, with an anomalously large correlation length. But while the probability distributions of various quantities show violations of finite size scaling,  we do not see the standard signs of an incipient first-order transition --- double-peaked probability distributions, etc. (Secs.~\ref{binder ratios section},~\ref{Pk section},~\ref{energy distribution section}). Fig.~\ref{heat capacity plot}  shows the heat capacity $C$, which quantifies the fluctuations of the energy. This has a diverging peak at large sizes, as expected for a critical point with a positive heat capacity exponent ($\alpha = 2- 3\nu>0$). The peak only emerges from the background at relatively large $L$. Surprisingly though, the peak fits well to a power law after subtracting a constant to account for the background,  $C_\text{max} -C_0\sim AL^{\alpha/\nu}$ (Fig.~\ref{heat capacity plot}, upper inset). This gives $\nu \simeq 0.44$, corresponding to a divergence $\sim L^{1.52}$. This divergence is much slower than the $L^3$ expected asymptotically at a first order transition. For a more intuitive picture, the lower panel of Fig.~\ref{heat capacity plot} shows the standard deviation of the energy, divided by the volume,  at $J=0.08850$. For a first order transition this should saturate to a constant (proportional to the square of the difference in energy density between the two phases) while here there is no sign of saturation. (More details in Sec.~\ref{energy distribution section}.)

To shed light on these perplexing observations, we analyse the topology of the RG flows in the $\nccp^{n-1}$ theory  in Sec.~\ref{RG flow section}. The topology we find allows for a scenario with an anomalously weak first-order transition for a range of $n$, as a result of  a coupling which `walks'  (runs slowly) in the proximity of a fixed point located at spatial dimension slightly below three. This is one possible reconciliation of the above numerical observations.  

A more radical possibility is that the transition is continuous but disobeys finite-size scaling because of a dangerously irrelevant variable. This was hypothesised for the $\mathrm{SU}(3)$ and $\mathrm{SU}(4)$ cases in Ref.~\cite{Kaul SU(3) SU(4)}. In this scenario we expect scaling violations in correlation functions when the separation $r$ of the points is comparable with $L$, but not when $r$ is fixed and $L\rightarrow\infty$. In Secs.~\ref{correlation functions section},~\ref{interpretation of results} we consider this possibility in the light of the data.

At present we cannot rule out either scenario; we sum up the situation in Sec.~\ref{interpretation of results}.

\section{Numerical results}
\label{numerical results section}

\subsection{N\'eel  and VBS order parameters \& correlators}
\label{order parameters}

\noindent
The deconfined transition separates phases that break different symmetries. In the VBS (short loop) phase, lattice symmetry is broken: this is quantified by the order parameter $\vec \varphi$ introduced in Sec.~\ref{loop model section}, whose spatially uniform part is 
\be\label{phi uniform part}
\vec \varphi = \f{\sqrt{2}}{N_\text{sites}} \lf \sum_{i\in A} \sigma_i , \sum_{i\in B} \sigma_j \ri.
\ee
This is normalised so $|\vec \varphi|^2 = 1$ for perfect VBS order (there are $N_\text{sites}/2$  sites on each sublattice). In the N\'eel (infinite loop) phase,  $\mathrm{SU}(2)$ spin-rotation symmetry is broken. In the loop representation, the magnitude of the N\'eel order parameter is the probability $N$ that a given link lies on an infinite loop \cite{cpn loops long}. For a finite system, one may define $N$ to be the probability that a link lies on a  strand that spans the system in the $z$ direction. If the transition is second order we expect finite-size scaling forms \cite{cardy book} for $\vec \varphi$ and $N$,
\ba\label{scaling forms - order parameters}
\langle |\vec \varphi |\rangle& = L^{-(1+\eta_\text{VBS})/2} f_{\varphi} \lf  L^{1/\nu} \delta J \ri, \\
N & = L^{-(1+\eta_{\text{N\'eel}} )/2} f_{N} \lf  L^{1/\nu} \delta J \ri, \label{scaling forms - order parameters2}
\end{align}
where $\delta J = J-J_c$. However attempting a scaling collapse using these forms gives negative $\eta$s and very poor collapse. Raw data for the order parameters was shown above in Fig.~\ref{orderparameters}.

\subsubsection{Correlation functions}
\label{correlation functions section}

\begin{figure}[t]
 \begin{center}
\includegraphics[width=\linewidth]{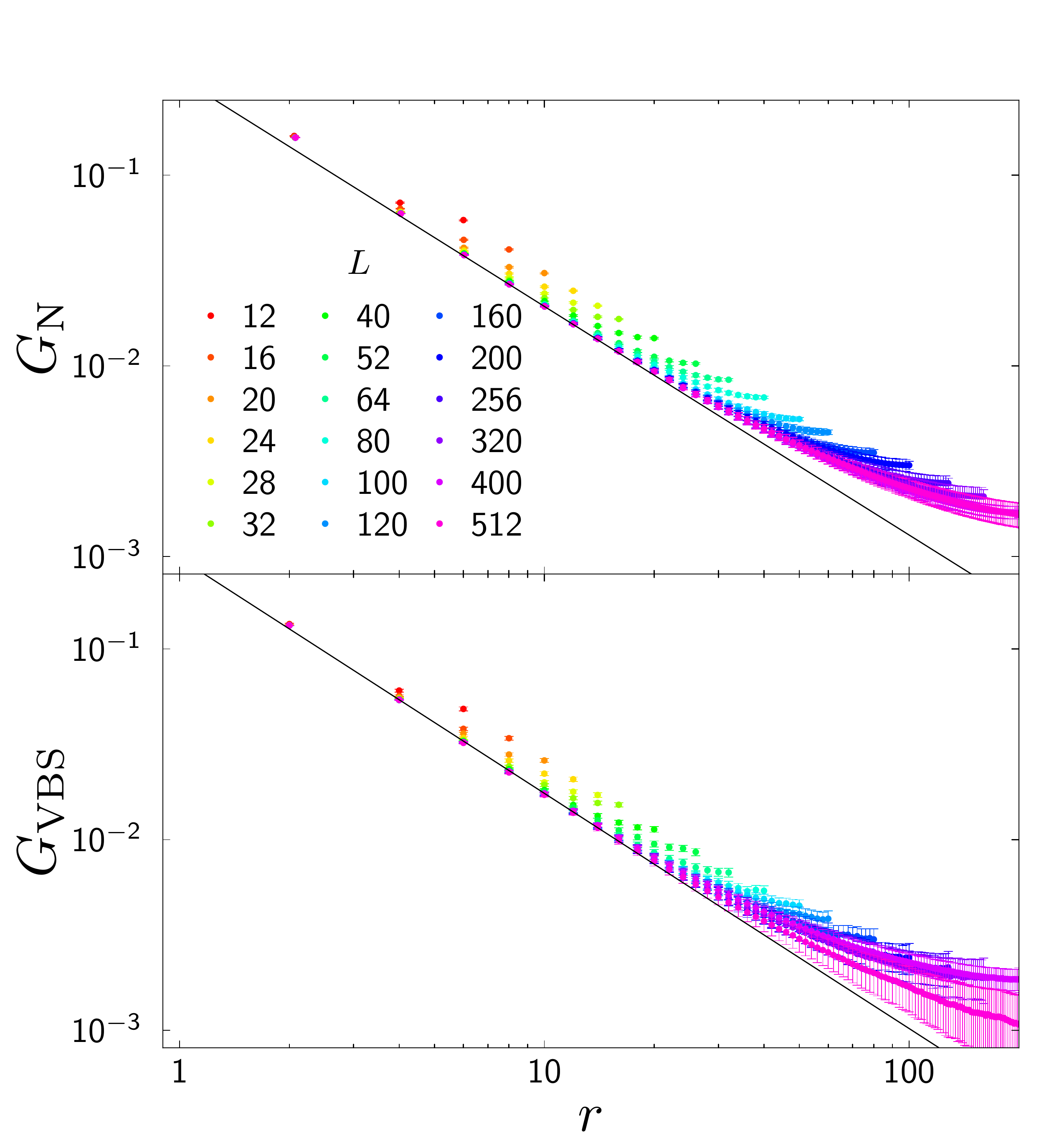}
 \end{center}
\caption{Correlators $G_\text{N\'eel}(r,L)$  (top) and $G_\text{VBS}(r,L)$ (bottom)  plotted against $r$, for various $L$.  Straight lines correspond to the estimates 
 { $\eta_\text{N\'eel} = 0.259(6)$ and $\eta_\text{VBS} = 0.25(3)$} from Eq.~\ref{eta estimates from Gprime}.
 \label{raw correlators}}
\end{figure}

\begin{figure} 
 \begin{center}
\includegraphics[width=\linewidth]{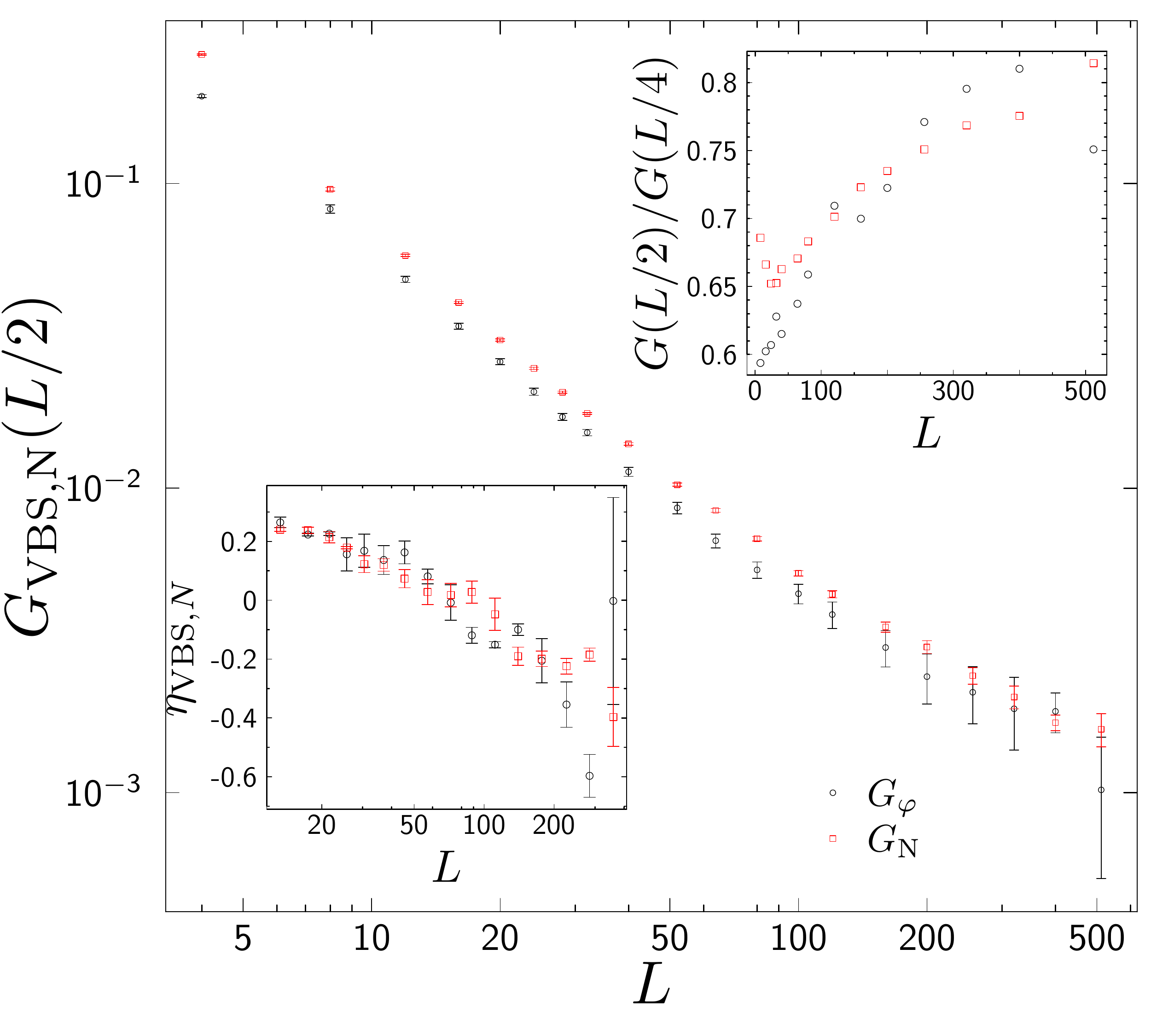}
 \end{center}
\caption{N\'eel and VBS correlation functions at separation $r=L/2$, plotted against $L$ on a log--log scale. The slopes give effective finite size exponent values $\eta_\text{N\'eel}(L)$ and $\eta_\text{VBS}(L)$, plotted in left inset. The right inset shows the ratio $G(L/2)/G(L/4)$ versus $L$. }\label{GL2correlator}
\end{figure}

\noindent
Next we examine the critical two-point correlation functions for $\vec \varphi$ and the N\'eel vector. In the loop representation, the N\'eel correlator is simply the probability that two links lie on the same loop \cite{cpn loops long}. We denote these correlators $G_\text{VBS}(r, L)$ and $G_\text{N\'eel}(r,L)$, where $r$ is the separation of the points (taken parallel to a coordinate axis) and $L$ is the system size. Raw data is show in Fig.~\ref{raw correlators}.

Conventionally, at $J_c$ one would expect
\be\label{conventional G ansatz}
G(r,L) = L^{-(1+\eta)} c(r/L),
\ee
with different $\eta$s and different scaling functions $c$ for each of the two observables. This would imply a collapse when plotting $L^{1+\eta} G(r,L)$ against $r/L$. Here this collapse fails, because the effective values of $\eta$ at small and at large distances differ, as we now quantify. 

The full correlation function is relatively complicated because it depends on two lengthscales, $r$ and $L$. Therefore a standard approach is to examine $G_\text{N\'eel}(L/2,L)$ and $G_\text{VBS}(L/2,L)$ as a function of $L$. According to (\ref{conventional G ansatz}) these scale as $L^{-(1+\eta_\text{N\'eel})}$ and $L^{-(1+\eta_\text{VBS})}$ respectively. In Fig.~\ref{GL2correlator} (main panel) we plot these correlators  against $L$ on a log--log scale. The gradual change of slope as a function of $L$ indicates a drift in the effective values of $\eta_\text{N\'eel}$ and $\eta_\text{VBS}$. The effective values $\eta_\text{N\'eel}(L)$, $\eta_\text{VBS}(L)$ determined from the slope are shown in  Fig.~\ref{GL2correlator} (lower inset). Note that for large $L$ the estimates for both exponents reach negative values. As mentioned above, negative values of $\eta_\text{VBS}$ or $\eta_\text{N\'eel}$ are ruled out for a continuous phase transition governed by a conformally invariant fixed point, though see below.

Another way to quantify the violation of finite-size scaling is via the ratios
\ba
&\f{G_{\text{VBS}}(L/2,L)}{G_{\text{VBS}}(L/4,L)}, &
&\f{G_{\text{N\'eel}}(L/2,L)}{G_{\text{N\'eel}}(L/4,L)},
\end{align}
which should be universal according to (\ref{conventional G ansatz}) but instead drift significantly with $L$; see Fig.~\ref{GL2correlator}, upper inset.

At certain critical points --- for example in $\phi^4$ theory above 4D --- a dangerously irrelevant variable invalidates standard finite-size scaling for the correlators. In this scenario it may happen that the correlator is conventional in the limit $L\rightarrow \infty$ (i.e. $G(r,\infty)\sim r^{-1-\eta}$ with $\eta\geq 0$), but anomalous when $r$ is comparable with $L$, or even with a smaller power of $L$ \cite{phi4 footnote}. Although  \textit{a priori} there is no theoretical reason to expect this phenomenon here, it suggests examining correlators in the regime $r\ll L$. From Fig.~\ref{raw correlators} it is conceivable that a well-defined power law will emerge in the limit $L\rightarrow \infty$, although  if so the convergence in $L$ is rather slow. 

\begin{figure} [t]
 \begin{center}
\includegraphics[width=\linewidth]{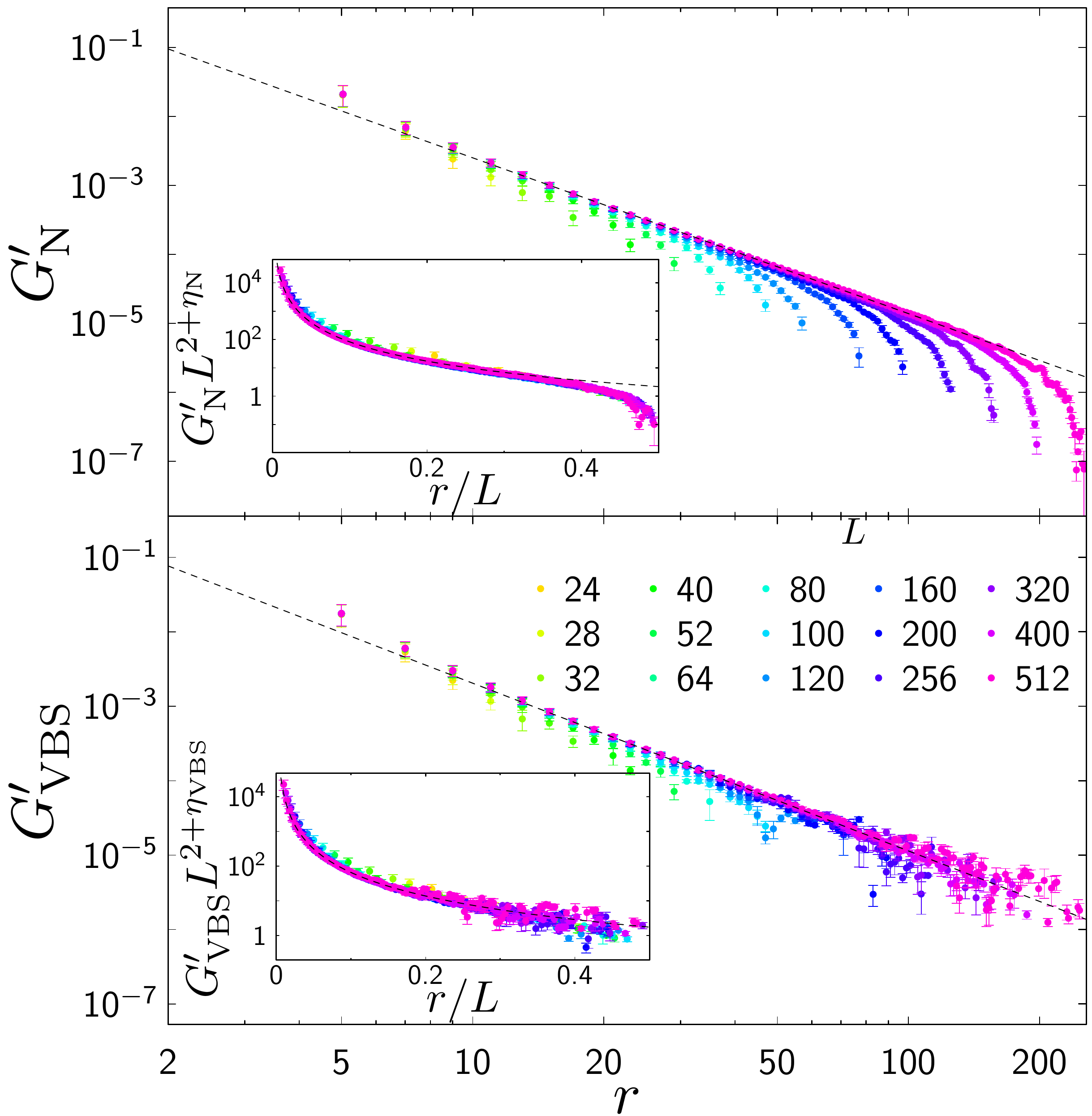}
 \end{center}
\caption{\emph{Derivatives} of N\'eel and VBS correlators, $\dd G_\text{N\'eel}(r)/\dd r$ and $\dd G_\text{VBS}(r)/\dd r$. Main panels show raw data and power-law fits (dashed lines) with { $\eta_\text{N\'eel} =0.259(6)$ and $\eta_\text{VBS} = 0.25(3)$.} Insets show scaling collapse of $L^{2+\eta} G'$ versus $r/L$, with the same values for  $\eta_\text{N\'eel, VBS}$. \label{Gderivatives}}
\end{figure}

For further insight we examine the \emph{derivatives} of the correlators, {$G'_\text{N\'eel}(r)=\dd G_\text{N\'eel}(r)/\dd r$} and {$G'_\text{VBS}(r)=\dd G_\text{VBS}(r)/\dd r$}, in Fig.~\ref{Gderivatives}. Remarkably, these quantities show quite clean power law behaviour up to at least $L\sim 100$ with
\ba\label{eta estimates from Gprime}
\eta_\text{N\'eel} & = { 0.259(6)} &
\eta_\text{VBS}& = { 0.25(3)},
\end{align}
 and quite good scaling collapse (insets to Fig.~\ref{Gderivatives}).  Reasonable scaling collapse is also obtained for `subtracted' correlators, defined as {$G(r,L) - G(L/2,L)$}. Straight lines corresponding to the above exponent values are shown in Fig.~\ref{raw correlators} for comparison with the raw data.

These results indicate that the strongest effect of the scaling violations is on the zero-modes of the fields, which have anomalously large fluctuations. This is suggestive because in $\phi^4$ theory above 4D,  the violation of finite-size scaling is caused by anomalously large fluctuations of the field's zero-mode \cite{brezin zinn justin fss}. The contribution of this mode to the two-point function depends on $L$ but not on $r$, so scaling can be repaired by differentiation/subtraction. The fact that this works perfectly in $\phi^4$ theory  is expected to be a special feature of the fixed point being free (and of the choice of correlator). Nevertheless, the good scaling of $G'_\text{N\'eel}$ and $G'_\text{VBS}$ is striking given the strong violation of scaling for the correlators themselves (Fig.~\ref{GL2correlator}), and may possibly indicate that a dangerously irrelevant variable is playing a role in the scaling violations.

\begin{figure}[t]
 \begin{center}
\includegraphics[width=0.95\linewidth]{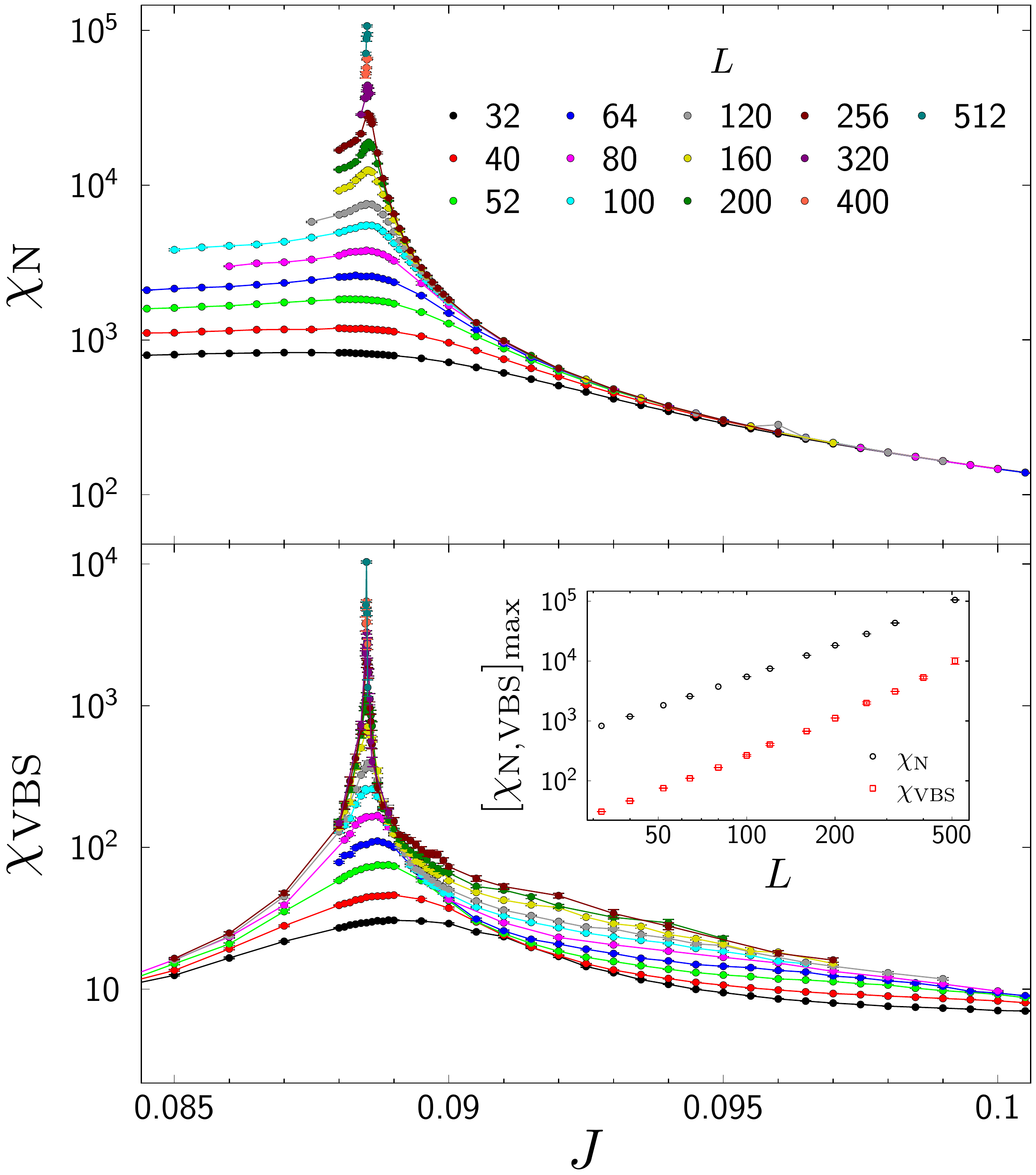}
 \end{center}
\caption{Susceptibilities for  N\'eel and VBS order parameters. Inset shows  peak heights as a function of $L$.
\label{susceptibilities}}
\end{figure}

\subsubsection{Fractal structure of loops}
\label{df section}

\noindent
The geometrical interpretation of the anomalous dimension $\eta_\text{N\'eel}$ is in terms of the fractal dimension of the loops, which according to conventional scaling relations is given by $d_f = (5-\eta_\text{N\'eel})/2$, and determines the power-law relation between the root mean square end-to-end distance $R$ of a strand and its length (see Ref.~\cite{cpn loops long} for details). This again gives a large positive $\eta_\text{N\'eel}$, in contrast to the drift towards negative values seen in the estimate from $G_\text{N\'eel}(L/2)$.  The simplest fit, taking  strands with $R\lesssim 100$ to minimise effects of finite $R/L$, gives $\eta_\text{N\'eel} = 0.42(6)$ (data not shown). We note that this is considerably larger than Eq.~\ref{eta estimates from Gprime}. However, attempting to include finite $R$ corrections in the fit gives smaller values in the range $0.25\lesssim \eta_\text{N\'eel}  \lesssim 0.42$ \cite{N versus Ns footnote}.

\subsubsection{Susceptibilities}
\label{susceptibility section}

\noindent
To compare with the estimates above, we calculate $\etan$ and $\etav$ from the N\'eel  \cite{cpn loops long}  and VBS \cite{VBS susceptibility} susceptibilities. These are shown in Fig.~\ref{susceptibilities}. According to finite-size scaling the peaks should diverge as $L^{2-\eta_\text{N\'eel,VBS}}$. The insets show log-log plots of the peak heights against $L$. The slopes indicate a downwards drift  from $\eta_\text{VBS}=0.164(13)$ to $\eta_\text{VBS}=-0.35(10)$, and  from $\eta_\text{N\'eel}=0.355(9)$ to $\eta_\text{N\'eel}=0.126(3)$.

\begin{figure}[t]
 \begin{center}
\includegraphics[width=\linewidth]{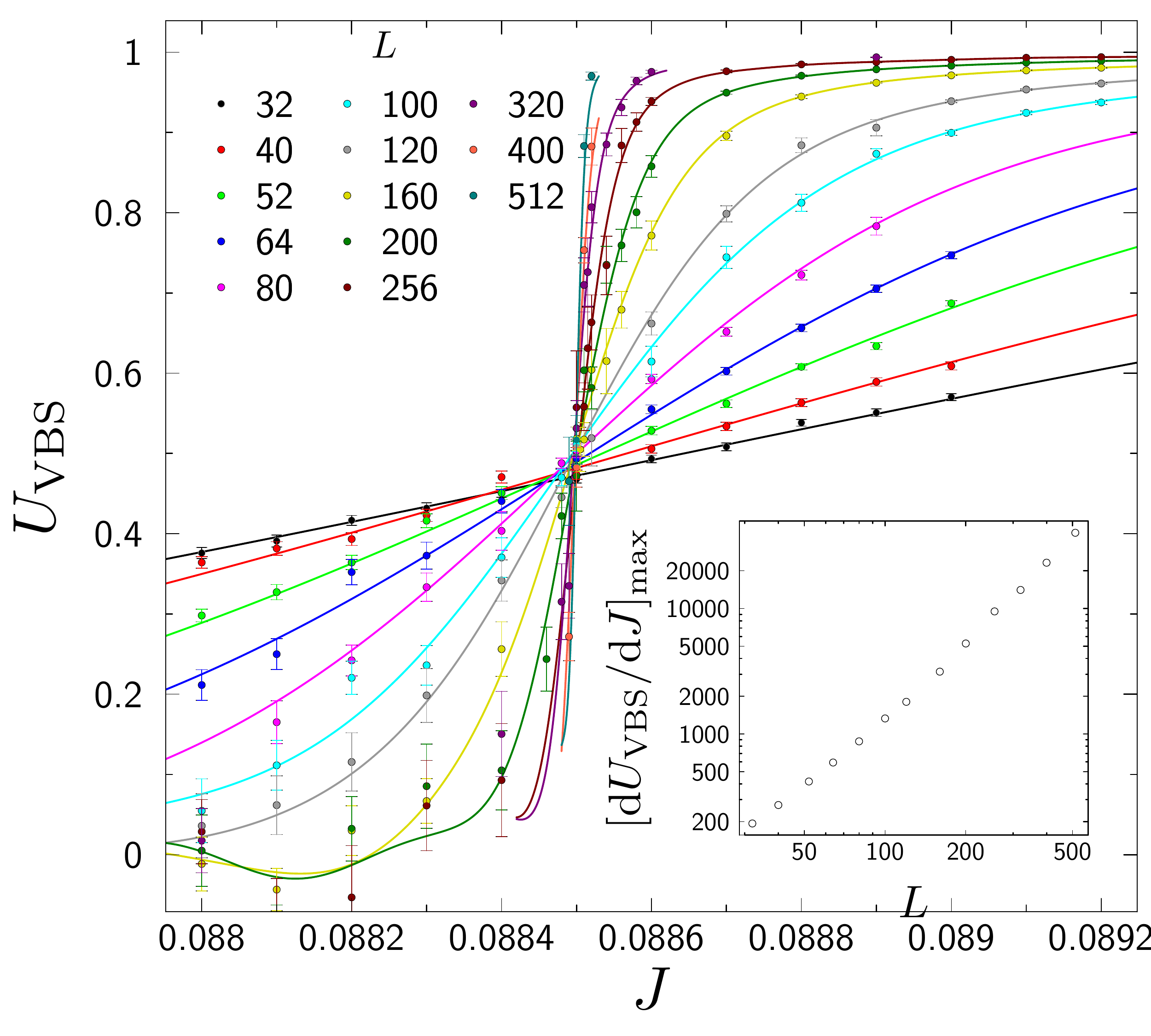}
  \end{center}
\caption{ $U_\text{VBS}$ versus $J$ for several system sizes ranging from 32 to 512. 
Inset: maximum value of $\de U_\text{VBS}/\de J$ versus $L$.\label{Bphi}}
\end{figure}

\subsubsection{Binder cumulant}
\label{binder ratios section}

\noindent
Fig.~\ref{Bphi} shows the Binder cumulant for the VBS order parameter, defined as
\begin{equation}
 U_\text{VBS} \equiv 2-\frac{\avrg{|\varphi|^4}}{\avrg{|\varphi|^2}^2}.
\label{binderB}\end{equation}
At a first order transition there should be a dip in $U_\text{VBS}$ which diverges with the system size \cite{Binder cumulant}. In our case there is no sign of this.

In the inset to Fig.~\ref{Bphi} we plot  the maximum value of the slope $\de U_\text{VBS}/ \de J$ for each $L$. For a second order transition this diverges as $L^{1/\nu}$ at the critical point. From the inset we see that there is different behaviour for small and large system sizes, giving $\nu=0.62(1)$ for sizes $L\le64$ and $\nu=0.476(18)$ for $L\ge256$. (See Sec.~\ref{corr exp drift} for other estimates of $\nu$.)

\subsection{Emergent symmetries}
\label{emergent symmetry section}

\begin{figure}[t]
 \begin{center}
 \includegraphics[width=0.9\linewidth]{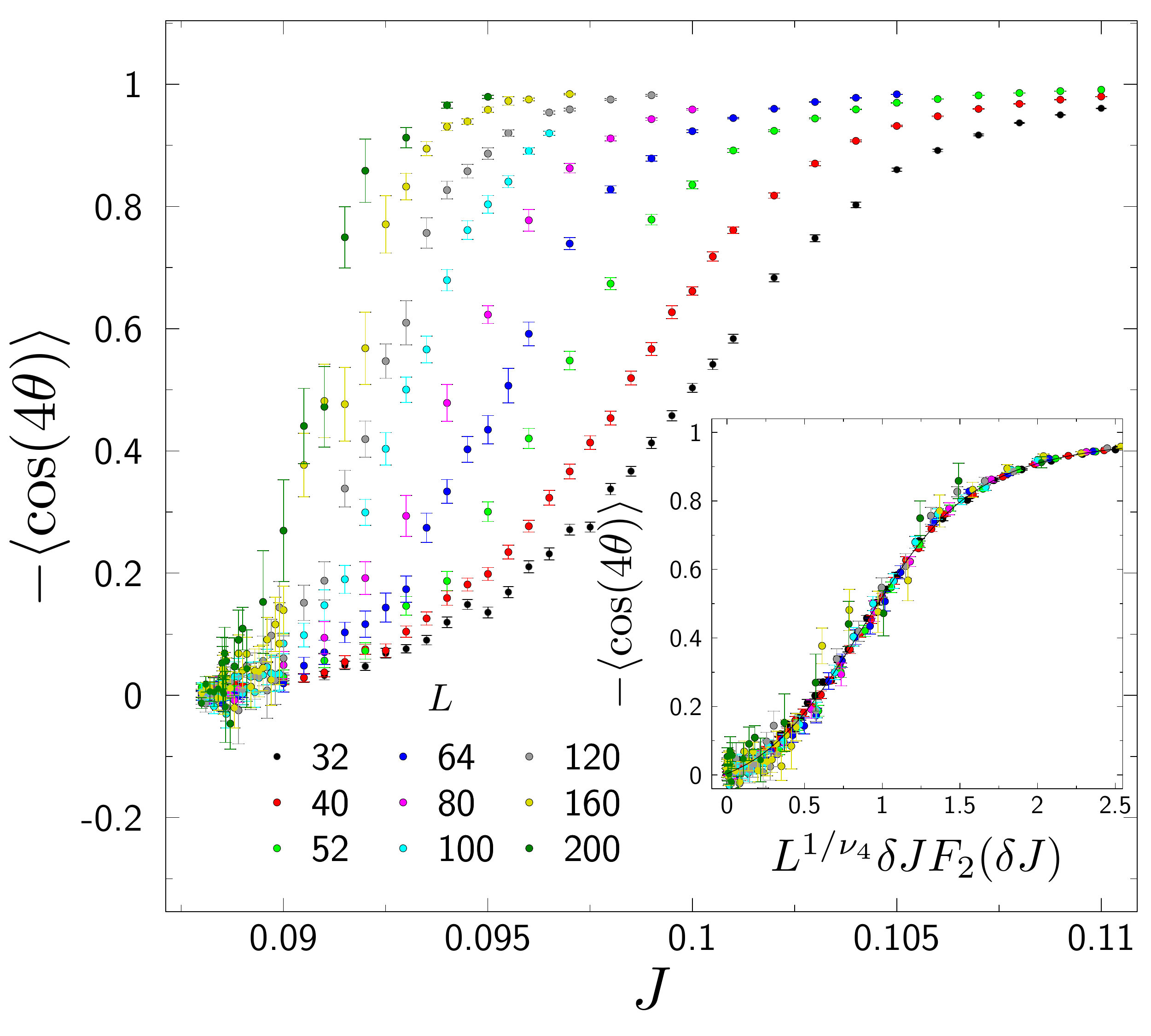}
 \end{center}
 \caption{Data for $\avrg{\cos(4\theta)}$ versus $J$ in several system sizes.
Inset: $\avrg{\cos(4\theta)}$ as a function of the scaling variable
$L^{1/\nu_4}(J-J_{\rm c})F_2(J-J_{\rm c})$. 
 \label{fig:C0}}
\end{figure}

\noindent
The deconfined criticality scenario assumes that the leading operator which breaks the symmetry for rotations of $\vec \varphi$ from $\mathrm{U}(1)$ down to $\mathbb{Z}_4$ is dangerously irrelevant: irrelevant at the critical point, but relevant within the VBS phase \cite{quantum criticality beyond}. This leads to the prediction of a crossover between $\mathrm{U}(1)$ and $\mathbb{Z}_4$ symmetry within the VBS phase,  on a lengthscale $\xi_\text{VBS}$ which is parametrically larger than the correlation length: $\xi_\text{VBS} \sim \xi^{1+|y_4|/3}$, where $y_4<0$ is the RG eigenvalue of the fourfold anisotropy \cite{Lou Sandvik Balents}. This has been confirmed in the J--Q model \cite{lou sandvik kawashima, block melko kaul}. 

Fig.~\ref{fig:2ddis} gave visual evidence for the emergent $\mathrm{U}(1)$ symmetry in the loop model. A quantitative measure of $\mathbb{Z}_4$ anisotropy is $\< \cos 4\theta\>$, where $\vec \varphi = |\vec \varphi| (\cos\theta,\sin\theta)$. Fig.~\ref{fig:C0} shows data  for sizes up to $L=200$. Ignoring scaling violations, the anisotropy should behave as \cite{Lou Sandvik Balents}
 \be
\< \cos 4 \theta \> = f\lf L^{1/\nu_4}\, \delta J \, F_2(\delta J)  \ri
\label{aniso}\ee
where $F_2(x)=1+ax+bx^2$ takes into account nonlinear dependence of the scaling variable on $J$ (needed here because of the larger range of $\delta J$ studied for this observable) and $\nu_4=\nu(1+|y_4|/3)$. The inset to Fig.~\ref{fig:C0} shows the attempted scaling collapse  using Eq.~\ref{aniso}. The exponent $\nu_4=1.09(6)$ is obtained from the fit. This confirms the irrelevance of fourfold anisotropy to the behaviour at the transition and to the explanation of the scaling violations. The corresponding value of $|y_4|$ is dependent on the assumed value of $\nu$, but is considerably larger than the estimate in Ref.~\cite{lou sandvik kawashima} for a variant of the J--Q model.

The closeness of the finite-size effective values of  $\eta_\text{VBS}$ and $\eta_\text{N\'eel}$ in Fig.~\ref{GL2correlator} and Eq.~\ref{eta estimates from Gprime} makes it tempting to speculate about a much larger emergent symmetry --- an $\mathrm{O}(5)$ symmetry relating the N\'eel and VBS vectors. This can be incorporated into an alternative field theory for the deconfined critical point \cite{senthil fisher competing orders} which was argued  to be equivalent to Eq.~\ref{NCCP1 lagrangian} \cite{O(5) footnote}. This symmetry enhancement would be analogous to the emergent $\mathrm{O}(4)$ symmetry of the 1D  spin-1/2 chain, which relates the spin-Peierls order parameter and the N\'eel vector \cite{1D chain footnote}. In the future it would be interesting to check explicitly for $\mathrm{O}(5)$ symmetry.

\subsection{N\'eel stiffness and  spanning strands}
\label{spanning number section}

\noindent 
A useful observable is the spanning number, $\mathcal{N}_s$,  defined as the number of strands that span the system in (say) the $z$ direction. Its mean value, $\<\mathcal{N}_s\>$, may be taken as a definition of the stiffness of the N\'eel order parameter. \cite{spanning number footnote} At a conventional critical point, $\<\mathcal{N}_s\>$ has scaling dimension zero and the scaling form
\be\label{spanning number scaling}
\<\mathcal{N}_s\> = h \big(  L^{1/\nu} \delta J \big).
\ee
Therefore $\<\mathcal{N}_s\>$ should be a universal constant at a critical point, modulo corrections due to irrelevant scaling variables: plots of $\<\mathcal{N}_s\>$ versus $J$ for different $L$ should cross at $J_c$. In the VBS phase $\<\mathcal{N}_s\>$ tends to zero exponentially in $L$, and in the N\'eel ordered phase it grows as $L$.

The mean spanning number was shown in Fig.~\ref{winding}. Contrary to the above expectation, $\<\mathcal{N}_s\>$ appears to diverge slowly with system size at the critical point. This is manifested in the upwards drift of the crossing points  in the main panel. In the inset we show pseudocritical values $\mathcal{N}_s^*(L)$ defined in two different ways. The data cannot be fitted with conventional scaling corrections from an irrelevant variable, i.e. $\mathcal{N}_s^* = \mathcal{N}_s^\text{crit} - A\, L^y$ with negative $y$: attempting such a fit leads to a positive (relevant) $y$. 

Previous work on the $\mathrm{SU}(2)$ J--Q model found a drift in a closely related winding number, and proposed that this indicated logarithmic corrections to scaling \cite{Sandvik logs}. Similar drifts were found for the $\mathrm{SU}(3)$ and $\mathrm{SU}(4)$ J--Q models \cite{Kaul SU(3) SU(4)}, fitting slightly better to a power law than a logarithm. The larger sizes considered here for the $\mathrm{SU}(2)$ case show that the divergence is certainly stronger than logarithmic. On attempting to represent it by a pure power law $\mathcal{N}_\mathrm{s}^*\sim AL^a$ we find that the exponent drifts upwards, but a power law plus constant
\ba
\mathcal{N}_\mathrm{s}^* & \simeq AL^{a} + B, 
& 
a& \simeq 0.6
\end{align}
fits the results for all $L$. This divergence is of course still slower than the linear behaviour expected asymptotically at a first order transition.

\subsubsection{Drift of critical probability distribution}
\label{Pk section}

\noindent
In addition to the mean $\<\spn\>$, we examine the full probability distribution of $\spn$.  Let $P_k$ be the probability that $\mathcal{N}_s$ is equal to $2k$, meaning that $k$ oriented strands span the system in a specified direction and $k$ in the reverse direction. This again has scaling dimension zero, so conventionally we would expect the scaling form
\be
\label{spanning number probability distribution}
P_k= g_{k} \big(  L^{1/\nu} \delta J \big).
\ee
By contrast, at a first order transition, where the short loop and infinite loop phases coexist, $P_k$ would have a peak at $k=0$  from the short loop phase and a peak at $k \propto L$  from the infinite loop phase.

\begin{figure}[t]
 \begin{center}
\includegraphics[width=0.9\linewidth]{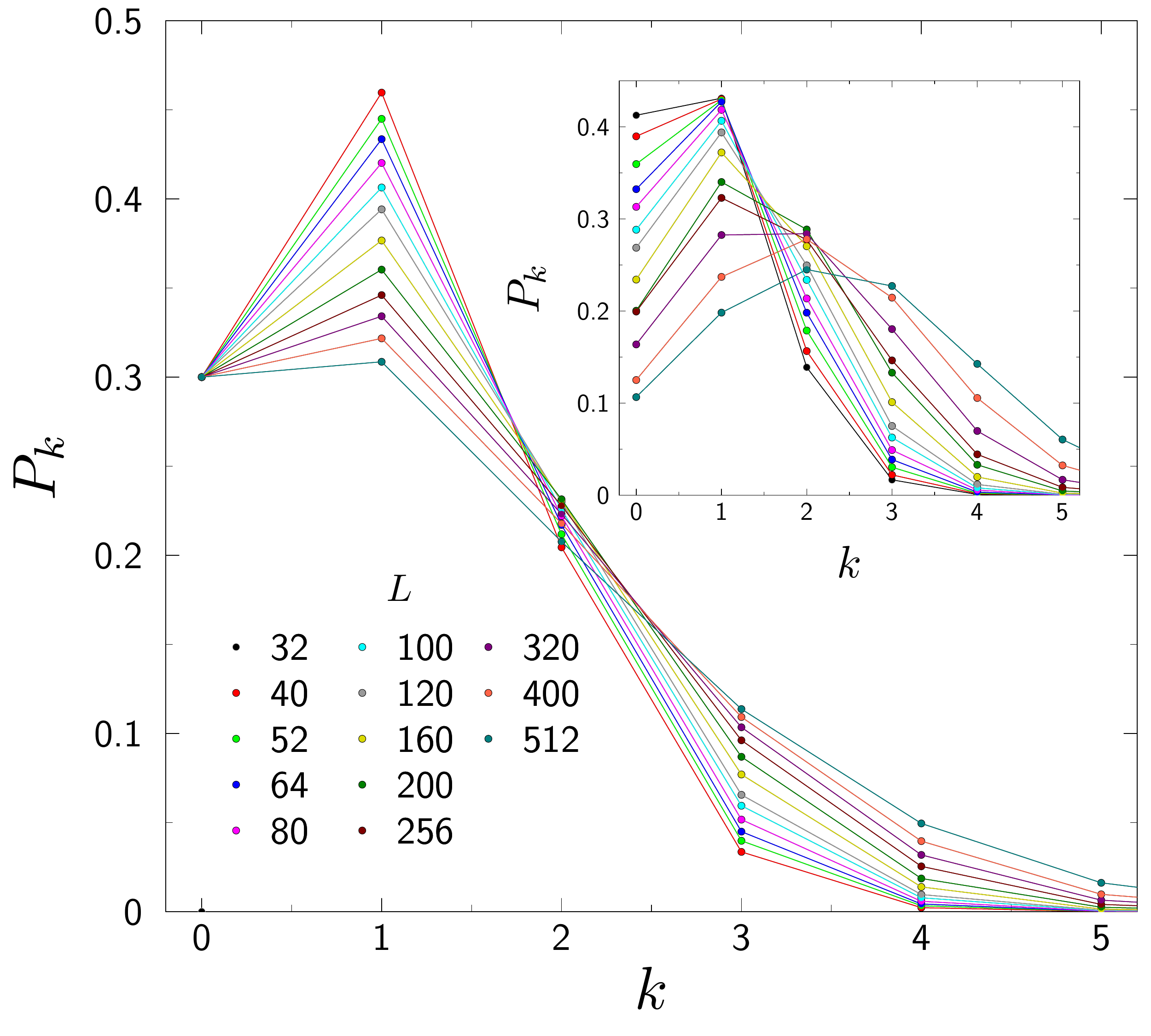}
 \end{center}
 \caption{Probability $P_k$ that the spanning number $\spn$ equals $2k$, for a range of $L$. In the main panel, $J$ is tuned so that $P_0=0.3$ for each $L$.  In the inset $J$ is fixed, $J=0.08850$.  \label{spanningprobdistr}}
\end{figure}

The distribution $P_k$ obtained numerically is shown in Fig.\ \ref{spanningprobdistr}, for various $L$. To compare different sizes, we tune $J$ for each $L$ so that $P_0 = 0.3$ (using the Ferrenberg method \cite{Ferrenberg}). For comparison, the inset shows the distribution at fixed $J = 0.0885$, very close to the critical point. Contrary to Eq.~\ref{spanning number probability distribution}, the data shows no sign of tending to a universal distribution. On the other hand, neither do we see a double-peaked structure developing.

\begin{figure}[h]
 \begin{center}
\includegraphics[width=0.95\linewidth]{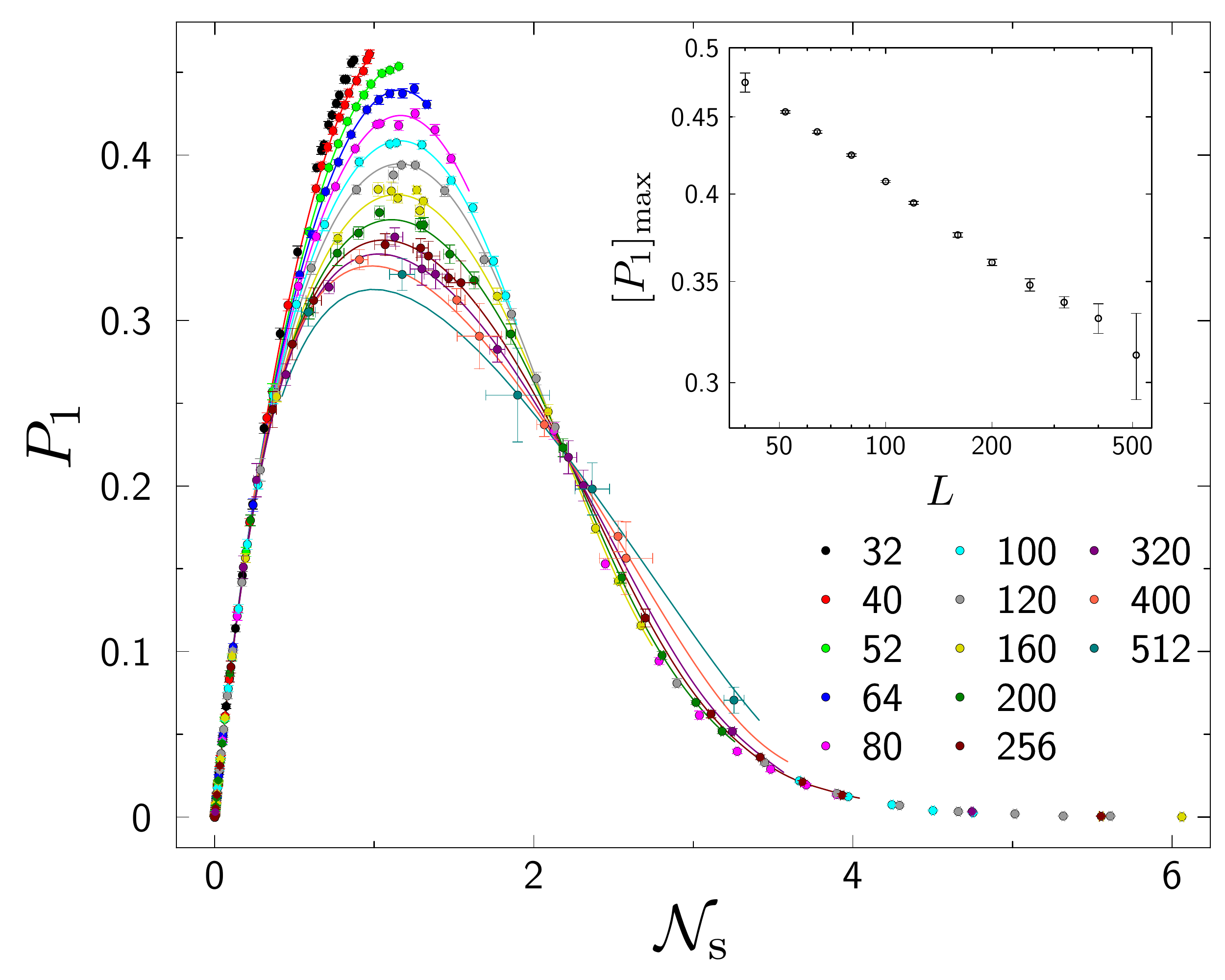}
 \end{center}
\caption{Parameter-free scaling:  $P_1(J)$  as a function of the average number ${\cal N}_{s}(J)$ of spaning curves.
 Inset: size dependence of   the maximum of  $P_1(J)$ on a double logarithmic scale. 
\label{P0}}
\end{figure}

The scaling form Eq.~\ref{spanning number probability distribution} would imply that $P_k$ is a universal function of $\<\mathcal{N}_s\>$ for each $k$. [Explicitly, $P_k = \lf g_k \circ h^{-1} \ri ( \<\mathcal{N}_s\> )$.] Therefore a plot of e.g. $P_1$ against $\<\mathcal{N}_s\>$ would show scaling collapse \emph{without the need to adjust any parameters}. (See Ref.~\cite{cpn loops long} for successful examples of such scaling collapse for the \emph{compact} $\cp^{1}$ [i.e. $O(3)$] and compact $\cp^2$ models.) It is  clear from Fig.~\ref{spanningprobdistr} that such a collapse will not work here. Fig.~\ref{P0} shows this for $P_1$ (interpolating curves are obtained with the Ferrenberg method \cite{Ferrenberg}). The dramatic failure to collapse is quantified in the inset, which shows the maximum value of $P_1$ as a function of $L$. At a conventional critical point this  would reach a finite constant as $L\rightarrow \infty$, while at a first order transition it should tend to zero. There is no evidence for saturation over this size range, though it cannot be ruled out. 

We note that the maximum of $P_1$ decreases below the universal value for the $\mathrm{O}(3)$ universality class \cite{cpn loops long}, which is close to $0.4$  (a stiffness in the J--Q model also drifts beyond the $\mathrm{O}(3)$ value
\cite{deconfined criticality flow JQ}). This is further confirmation that we are dealing with a direct transition rather than two separate transitions that are too close to be resolved.

\subsection{Energy distribution}
\label{energy distribution section}

\begin{figure}[t]
 \begin{center}
\includegraphics[width=0.8\linewidth]{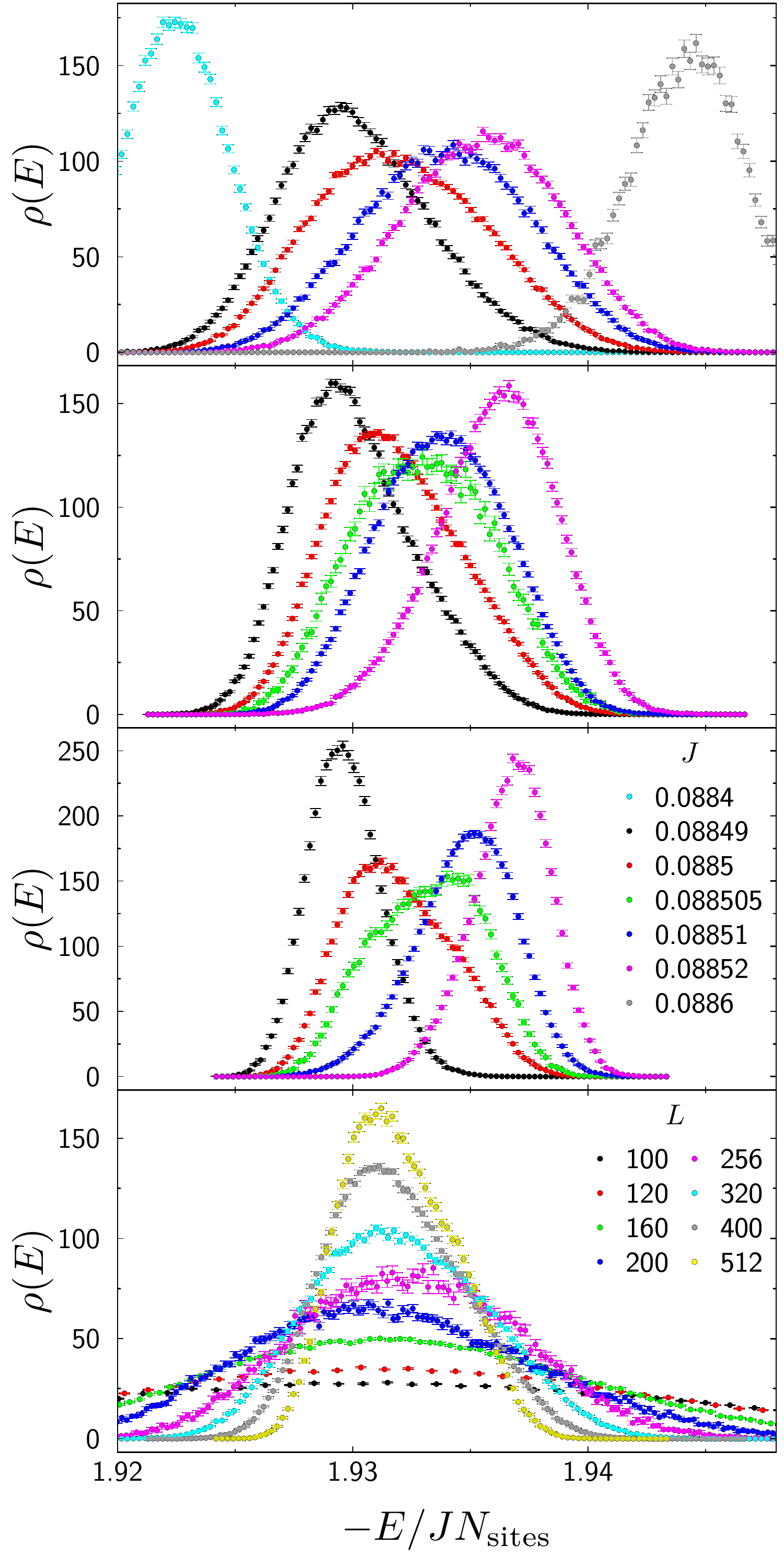}
 \end{center}
\caption{First three panels starting at top: energy distribution functions for system sizes $L=320$, $400$ and $512$ respectively, for various $J$. Bottom panel: energy distribution for $J=0.08850$ for various $L$.\label{energy}}
\end{figure}

\begin{figure}[t]
 \begin{center}
\includegraphics[width=\linewidth]{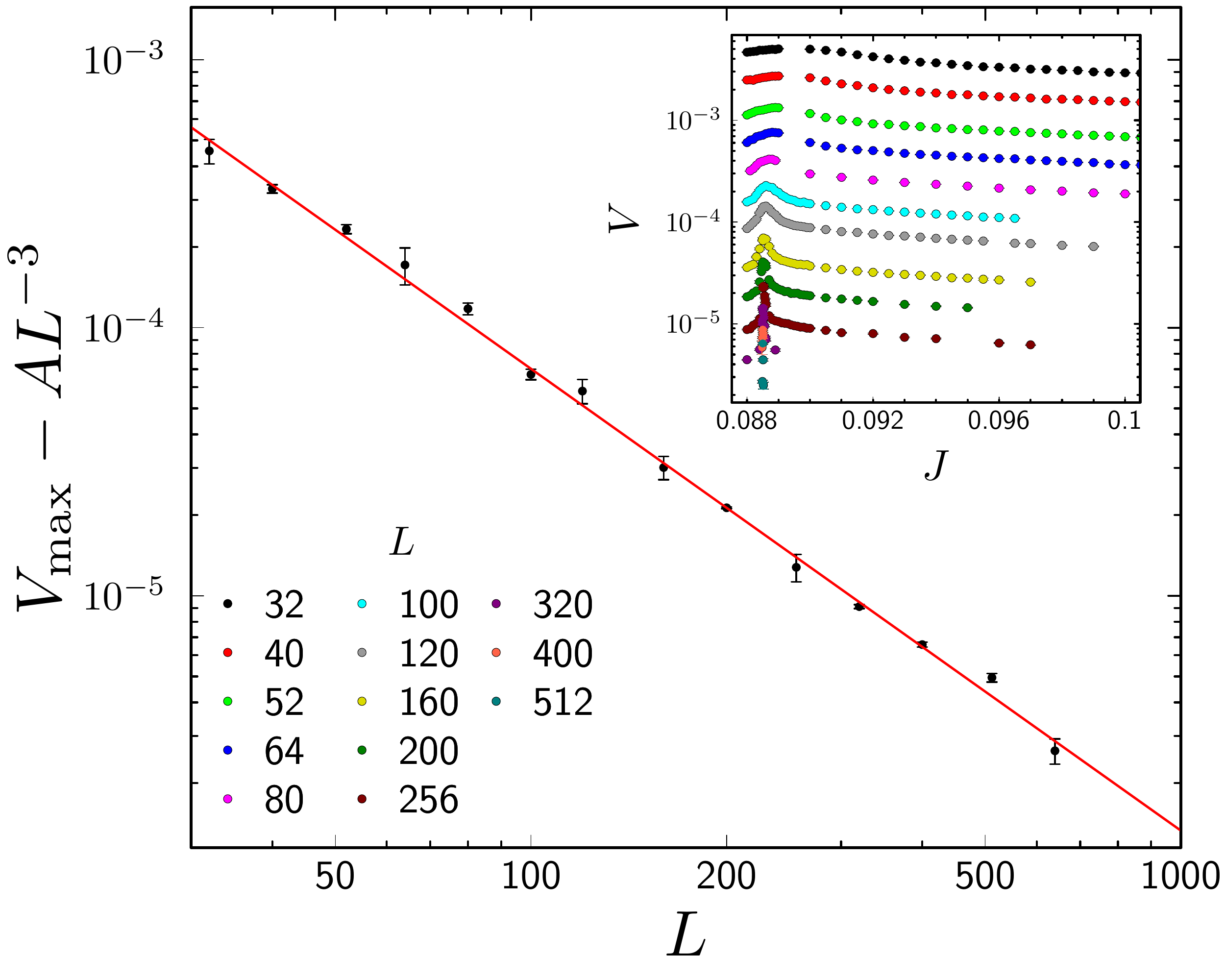}
 \end{center}
\caption{$V_{\rm max}$ is shown as a function of system size on a double logarithmic scale with the background contribution substracted. Inset: raw data for $V$ plotted as a function of $J$ on a semi-logarithmic scale for several values of $L$. \label{VL}}
\end{figure}

\noindent
The behaviour of the heat capacity, proportional to the variance in the energy $E$, has been discussed in connection with Fig. 6. Further information is contained in the full probability distribution for $E$ (which is defined in Eq.~\ref{full partition function}). The top panels of Fig.~\ref{energy} show how this distribution evolves as $J$ is varied. The bottom panel shows the distribution at $J\simeq J_c$ for various $L$. We do not see a double-peaked distribution. The width of the critical distribution also decreases with increasing system size, contrary to the expectation for a first order transition (Fig.~\ref{heat capacity plot}, lower inset).

The Binder parameter
\begin{equation}
 V= \avrg{E^4} /{\avrg{E^2}}^2-1,
\label{binder}\end{equation}
plotted in the inset to Fig.~\ref{VL}, is an alternative quantity for analysis.  The data shows a peak near the transition, with a  height $V_\text{max}$ that decreases with $L$ (it is necessary to use the Ferrenberg interpolation method \cite{Ferrenberg} for accurate estimates of $V_\text{max}$).  At a first-order transition $V_\text{max}$ should saturate to a constant as a result of the double-peaked energy distribution, while at a continuous transition with $\alpha >0$ the peak height should tend to zero as $V_\text{max} \sim L^{2/\nu-6}$. Here, a direct estimate of $\nu$ using the slope gives a value that drifts from $\nu\sim0.621(5)$ for system sizes $32\le L\le 64$ to $\nu\sim0.481(12)$ for $L\ge256$.  However, in addition to the peak, $V$ has a large background contribution scaling as $L^{-3}$ (see inset). It is natural to subtract such a correction. This allows $V_\text{max}$ to be fitted to the the power law form corresponding to  $\nu=0.468(6)$  --- see Fig.~\ref{VL}, main panel.

\subsection{Correlation length exponent}
\label{corr exp drift}

 \begin{figure}[t]
 \begin{center}
\includegraphics[width=0.95\linewidth]{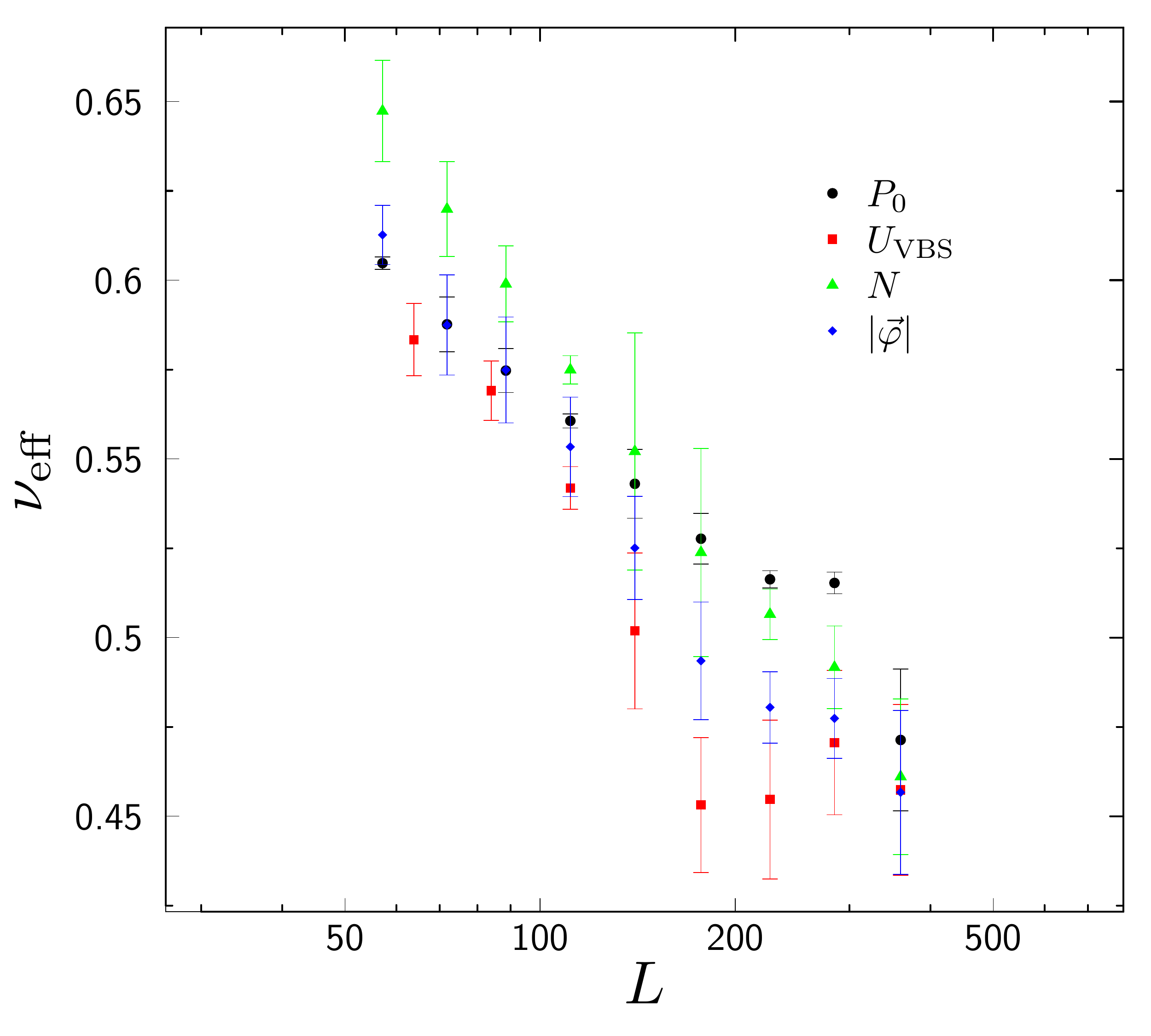}
 \end{center}
\caption{ Size dependence of the effective  correlation length exponent $\nu_\text{eff}$  obtained from finite-size scaling of various quantities. 
\label{effectivenu}}
\end{figure}

 \begin{figure}[t]
 \begin{center}
\includegraphics[width=0.9\linewidth]{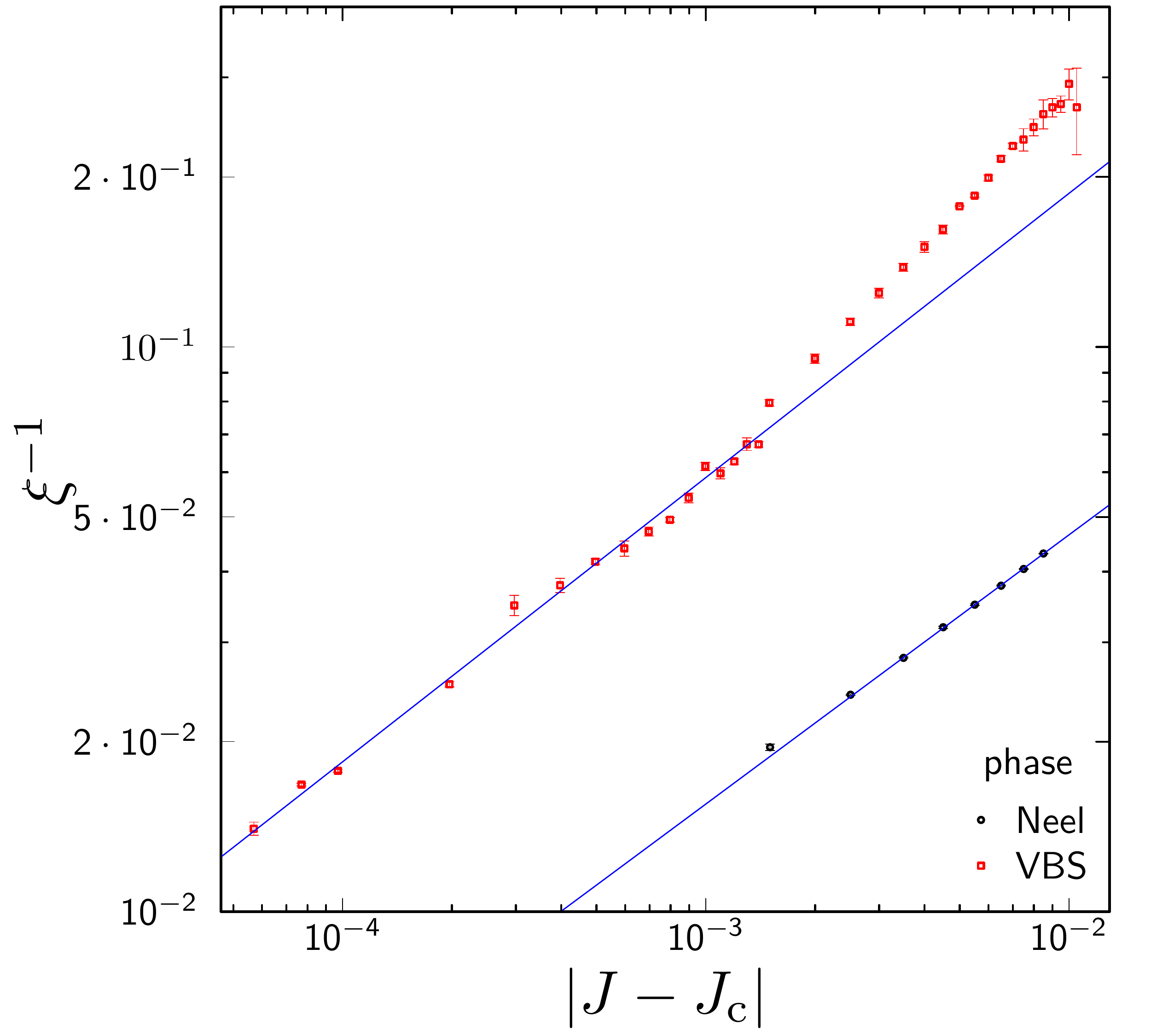}
 \end{center}
\caption{Estimates of the correlation length obtained from (1) the exponential decay of the spanning number $\spn$ in the VBS phase, $J>J_c$ (red points); (2) the coefficient of the linear growth of $\spn$ in the N\'eel phase, $J<J_c$ (black points) Note: the overall normalisation is arbitrary in the latter case. \label{corrlengthfig}}
\end{figure}

\noindent
In this section we discuss two different approaches to determining the correlation length exponent $\nu$, one relying on finite size scaling and one not  (see Sec.~\ref{order parameters} for anomalous dimensions $\etan$, $\etav$).  

First, we obtain estimates using the standard finite-size scaling forms for various observables. For example  the Binder cumulant for the VBS order parameter $\vec\varphi$ would naively scale as $U_\text{VBS}=f_U(L^{1/\nu}\delta J)$, so that the maximum value of $\dd U_\text{VBS}/\dd J$ should grow as $L^{1/\nu}$. Therefore we can define an effective exponent via $\nu_{\rm eff}(L)^ {-1}= {\rm d}\log([U'_\text{VBS}]_{\rm max})/{d \log L}$. We calculate such numerical derivatives using four consecutive system sizes.

Fig.~\ref{effectivenu} shows  the resulting estimates $\nu_{\rm eff}(L)$  from $U_\text{VBS}$, from the probability $P_0$ of having no spanning strands, and the order parameters $N$ and $|\vec  \varphi   |$.  $\nu_{\rm eff}(L)$  drifts from large values, around 0.62 (in accordance with previous studies \cite{Sandvik JQ, melko kaul fan, Banerjee et al, lou sandvik kawashima, Kawashima deconfined criticality}) to values around 0.46. The latter is in agreement with the estimate from the heat capacity with the background subtracted (Sec.~\ref{numerics overview}).  Despite the drift, $\nu_\text{eff}(L)$ remains above the unitarity bound 2/5.  A drift in $\nu$ was also identified in Ref.~\cite{Kawashima deconfined criticality}.

The above estimates all rely on finite-size scaling forms for observables on the scale of the system size. To avoid the assumption of conventional finite-size scaling, we also estimate $\nu$ directly from the correlation length $\xi$ in the regime where $\xi\ll L$. For values of $J$ in the VBS (short loop) phase, we determine $\xi$ by fitting the $L$-dependence of the spanning number to the expected form $\spn \propto L^2 e^{-L/\xi(J)}$. In the N\'eel phase, the spanning number is expected to grow as $\spn \sim A L/\xi(J)$. This allows us to determine $\xi(J)$ up to the overall constant $A$.

The results are shown in Fig.~\ref{corrlengthfig}. The power-law fits shown give $\nu=0.477(4)$ for the data in the N\'eel phase and $\nu = 0.503(9)$ for the data in the VBS phase. These values are close to the estimates in Fig.~\ref{effectivenu} at the largest sizes. But it is remarkable that here this behaviour sets in at \emph{much} smaller lengthscales. For the VBS phase (where we can determine $\xi(J)$ without the complication of an overall constant), the above exponent fits the data well starting from scales as small as $\xi \sim 15$.

We believe that if the transition is indeed continuous, the correlation length exponent is close to $\nu =0.5$, so considerably smaller than most earlier estimates from J--Q and $\nccp^1$ models. This value is close to $2+\epsilon$ expansion results for the $\cp^1$ nonlinear sigma model which should apply to the deconfined critical point assuming the transition is continuous (see Sec.~\ref{failure of 2+epsilon}).

\section{Field theory for loop model; hedgehog fugacities}
\label{continuum description section}

\noindent
Models for completely-packed, oriented loops can be mapped to lattice $\cp^{n-1}$ models with an unconventional but simple form \cite{cpn loops short, cpn loops long}. At first sight the continuum limit of these models is simply the (compact) $\cp^{n-1}$ sigma model. Here we discuss this continuum limit in more detail, and show that hedgehog defects in the $\cp^{n-1}$ spin configuration contribute imaginary terms to the action that are analogous to the Berry phases in the Euclidean action for the 2D quantum Heisenberg model \cite{Haldane 2+1, Read Sachdev VBS and spin peierls}. These terms are crucial for the present model. By the reasoning of Refs.~\cite{deconfined critical points, quantum criticality beyond, critically defined, Motrunich Vishwanath}, they change the  effective continuum description from the compact $\cp^{n-1}$ model to $\nccp^{n-1}$. (By contrast, the imaginary terms were unimportant for the transitions discussed in Refs.~\cite{cpn loops short, cpn loops long} as a result of the lower lattice symmetry there.)

The quantities of interest are determined by symmetry, so it is enough to consider the case $J=0$, where the lattice field theory for the loop model is simplest. The $\cp^{n-1}$ spins are placed on the links $l$ of the lattice. They are complex vectors $\zb_l = (z_l^1, \ldots, z_l^n)$, with fixed length $|\zb|^2 = 1$ and the gauge redundancy $\zb_l \sim e^{i\varphi_l} \zb_l$. In a loose notation where the incoming links at a given node are denoted $i$ and $i'$ and the outgoing links $o$ and $o'$, the partition function is
\be \label{lattice cpn-1 model}
Z = \Tr \prod_\text{nodes} \lf \f{1}{2} (\zb_o^\dag \zb_i) (\zb_{o'}^\dag \zb_{i'}) + \f{1}{2} (\zb_{o'}^\dag \zb_i) (\zb_{o}^\dag \zb_{i'})  \ri.
\ee
Here `$\Tr$' is the integral over the $\zb$s. Under a gauge transformation of $\zb_l$, the terms for the two nodes adjacent to $l$ pick up opposite phases, so the Boltzmann weight is invariant. The mapping between (\ref{lattice cpn-1 model}) and the loop model follows from a straightforward graphical expansion which is described in \cite{cpn loops long}.

Let us consider the continuum description of (\ref{lattice cpn-1 model}). To begin with, take a configuration in which $\zb$ is slowly varying. Each term in the product over nodes is then close to one, and we may obtain a continuum sigma model Lagrangian by a derivative expansion. In 3D, the only term with two derivatives allowed by global, gauge and lattice symmetries is the standard sigma model kinetic term. Let us focus on the $n=2$ case, and parameterise $\cp^1$ (which is simply the sphere) using the N\'eel vector
\ba
N_a &= {\zb}^\dag \sigma_a \zb, & a&=x,\, y,\, z,
\end{align}
instead of the gauge-redundant field $\zb$. Then
\ba\label{O3 lagrangian}
\mathcal{L}_\sigma &= \f{K}{2}(\nabla \vec N)^2, & &(\vec N^2 = 1).
\end{align} 
A crude way to estimate a bare value of $K$ is to calculate the Boltzmann weight in (\ref{lattice cpn-1 model}) for a spin configuration with a uniform twist, giving  $K=1/16$. (For general $n$, $\lag_\sigma$ may be written in terms of the matrix $Q = \zb \zb^\dag - \mathbb{1}/n$.)

The Lagrangian obtained by the derivative expansion can fail to capture the true scaling behaviour in two ways. It fails in a trivial way when $\vec N$ varies strongly at a node. This will of course be the case in the lattice model, and leads to an order one renormalisation of  the stiffness.

More importantly, the \emph{phase} of $\zb$ can vary rapidly even if $\vec N$ is slowly varying. Nodes where $\vec N$ is approximately constant but where this phase varies abruptly contribute imaginary terms to the action. For smooth configurations of trivial topology, these phases cancel. However, in the presence of hedgehog defects, there remain nontrivial phases that are missed by the derivative expansion.

This is because in a configuration with a hedgehog it is \emph{impossible} to find a gauge in which $\zb$ is everywhere slowly varying, even far from the hedgehog core. This follows from the fact that topological flux density $B_\mu$, which when integrated over a closed surface gives the signed number of hedgehogs inside, is a total derivative when written in terms of $\zb$: $B_\mu = \f{1}{i} \epsilon_{\mu\nu\lambda} \nabla_\mu (\zb^\dag \nabla_\nu \zb)$. If $\zb$ were continuous, integrating the topological density over a large sphere would give zero. Therefore $\zb$ must be discontinuous somewhere on the sphere if the sphere encloses a hedgehog.

A simple calculation is required to determine what effect the imaginary terms have on the weight for a configuration with a hedgehog. We do this calculation in Appendix~\ref{hedgehog calculation}. For specificity, we take the hedgehog to be centred on a site of $C_1$ or a site of $C_2$ --- for example at the centre of the cube in Fig.~\ref{L lattice figure}. These locations form a bcc lattice, with four sublattices. We find that the weight of a configuration with a hedgehog acquires a fugacity proportional to 
\be\label{values of hedgehog fugacity}
1,\quad i, \quad -1 \quad \text{or} \quad -i, 
\ee
depending on which sublattice it sits on. (More precisely, only the relative phase between different locations is meaningful \cite{relative phase footnote}.)

This also generalises immediately to larger $n$. The result matches nicely what is found for the 2D quantum Heisenberg model \cite{Haldane 2+1, Read Sachdev VBS and spin peierls, Read Sachdev VBS and spin peierls 2}, where the  fugacity for instantons --- hedgehogs in spacetime --- takes the same set of values as above depending on which of four sublattices of the square lattice the instanton occurs on.  

By symmetry, we infer that the coarse-grained hedgehog fugacity vanishes. That is, it vanishes as a result of phase cancellation between configurations in which the hedgehog is centred on nearby sites on different sublattices. Thus the arguments of Refs.~\cite{deconfined critical points, quantum criticality beyond, critically defined, Motrunich Vishwanath} apply, giving the $\nccp^1$ model (Eq.~\ref{NCCP1 lagrangian}) as the continuum description. 

An alternative argument for the $\nccp^1$ description of square lattice spin-1/2 antiferromagnets was given in Ref.~\cite{levin senthil}, focusing on the VBS order parameter $\vec\varphi$ and its vortex defects rather than the N\'eel order parameter $\vec N$ and its hedgehog defects. The key point of this alternative argument is that a vortex in the VBS order parameter carries a single unpaired spin at its centre. In spacetime, this corresponds to an extended spinon worldline running along the vortex core. This has a direct interpretation in the loop model: a vortex line in the node order parameter $\vec \varphi$ has a single extended loop running along its core. We have confirmed this explicitly by constructing such configurations.

In previous work we have considered transitions in a different version of the loop model which does \emph{not} preserve the full lattice symmetry \cite{cpn loops short, cpn loops long}. The transitions in that less-symmetric model are described by the \emph{compact} $\cp^{n-1}$ model, unlike the present loop model whose  transition is described by $\nccp^{n-1}$. This is because breaking lattice symmetry spoils the cancellation between the values in Eq.~\ref{values of hedgehog fugacity}, leaving a nonzero hedgehog fugacity. This allows the standard critical behaviour of the compact $\cp^{n-1}$ model (i.e. of the usual $\mathrm{O}(3)$ model when $n=2$).  We note that in the case $n=3$, the \emph{compact} $\cp^2$ model appears to show an interesting continuous transition which is naively forbidden by Landau theory \cite{cpn loops short, cpn loops long, kaul-su(n)bilayer}.  An RG explanation for why the expectation from Landau theory  breaks down in this case was given in Ref.~\cite{cpn loops long}.

\section{RG flows in the $\nccp^{n-1}$ model \\ ($n$--component Abelian Higgs model)}
\label{RG flow section}

\noindent
In this section we make a conjecture for the topology of the RG flows in the $\nccp^{n-1}$ model, the $n$-component generalisation of Eq.~\ref{NCCP1 lagrangian} with $\zb =(z_1, \ldots, z_n)$:
\be\label{NCCPn-1 lagrangian}
\mathcal{L} = |(\nabla - i A) \zb|^2 + \kappa (\nabla \times A)^2 + \mu |\zb|^2 + \lambda |\zb|^4.
\ee
This is also known as the $n$-component Abelian Higgs model. We will treat both $n$ and the spatial dimension $d$  (between 2 and 4) as continuously varying.

Scaling violations are seen in a wide variety of different lattice models that are related to this field theory (at $n=2$) and persist to very large lengthscales \cite{unimportance of fourfold anisotropy footnote, dimer model footnote}, so we believe a plausible explanation for them should appeal to  universal physics of the $\nccp^{n-1}$ model and not to accidental features of specific Hamiltonians. Results for $\mathrm{SU}(3)$ and $\mathrm{SU}(4)$-symmetric models ($n=3, \,4$ \cite{Kaul SU(3) SU(4), Kawashima deconfined criticality}) suggest that a satisfactory explanation should also account for scaling variations across a range of $n$. 

\begin{figure}[t]
 \begin{center}
 \includegraphics[width=\linewidth]{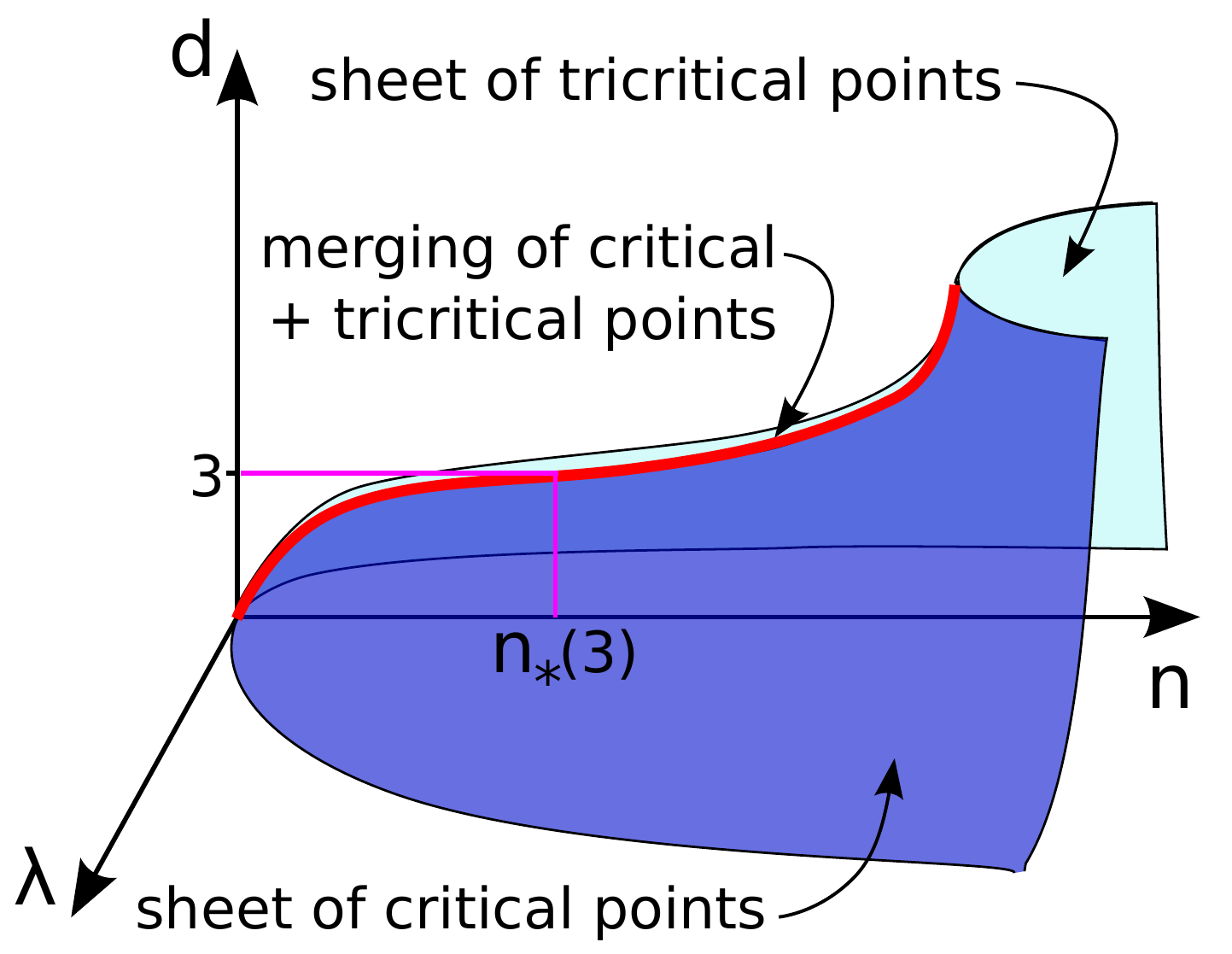}
 \end{center}
 \caption{Figure showing topology of the sheet of RG fixed points for dimensionalities around three (the limiting cases $d=2$ and $d=4$ are not shown).}
 \label{rgflows3d}
\end{figure}

Fig.~\ref{rgflows3d} shows the basic topology of the RG flows we find. This is a sheet of  RG fixed points projected on the space of $n$, $d$, and a scaling variable $\lambda$. (Close to 4D, $\lambda$ is the quartic coupling in Eq.~\ref{NCCP1 lagrangian}, but in lower dimensions one cannot make this identification.) The RG flow is parallel/antiparallel to the $\lambda$ axis, since $n$ and $d$ do not flow. $\lambda$ is irrelevant on the critical sheet and relevant on the tricritical sheet. The strongly relevant coupling which drives the transition (i.e. the mass) is not shown, since we consider the theory at the critical value. For any fixed value of $n$ between zero and $n_{*4}\simeq 183$, the critical point exists so long as we are sufficiently close to two dimensions. When $d$ is increased, the critical point disappears at a universal value $d_*(n)$, by merging with the tricritical point.

To begin with, consider three limits in which the $\nccp^{n-1}$ model is solvable. First, it  is tractable by saddle-point at large $n$, where it yields a nontrivial critical point for $2<d<4$. This critical point describes a direct transition between a Higgs phase, where $\zb$ is condensed and $\mathrm{SU}(n)$ symmetry is broken, and a Coulomb phase where $\mathrm{SU}(n)$ symmetry is unbroken and the gauge field $A$ is massless.

The field theory is also tractable in a $4-\epsilon$ expansion~\cite{halperin lubensky ma}. For infinitesimal $\epsilon$, a weak--coupling critical point exists only if $n$ is greater than or equal to a value which we denote $n_{*4}$. This value is quite large, $ n_{*4}\simeq 183$. In fact in the regime $n>n_{*4}$ where the critical point exists, the $4-\epsilon$ expansion also yields a \emph{tricritical} point at a smaller value of  the quartic coupling $\lambda$. As $n$ approaches $n_{*4}$ from above, these two fixed points approach each other, and they annihilate  when $n$ reaches $n_{*4}$. For $n<n_{*4}$ there is no nontrivial fixed point: the theory is expected to flow to a discontinuity fixed point at large negative $\lambda$ representing a first order transition.

Finally, the $\nccp^{n-1}$ model can be studied in $2+\epsilon$ dimensions by switching from a soft-spin formulation to a nonlinear sigma model \cite{hikami beta functions 1981, lawrie athorne}. In this regime a continuous phase transition is found for all values of $n$ greater than zero. (The `replica-like' regime  $n\leq 1$ is meaningful and describes certain classical loop models \cite{cpn loops short, loops with crossings, vortex lines}.) In the $2+\epsilon$ approach it does not matter whether the nonlinear sigma model is formulated  with a dynamical  $\mathrm{U}(1)$ gauge field, or as a pure nonlinear sigma model with target space $\cp^{n-1}$: the two formulations give identical results \cite{lawrie athorne}. (When the dynamical gauge field is included its coupling flows to infinity, so  it can be integrated out leaving the usual  $\cp^{n-1}$ nonlinear sigma model.)

A crucial point, made  in Ref.~\cite{March Russell}, is that the fixed point found in the large $n$ approach is the \emph{same} as that found in both the $2+\epsilon$ and $4-\epsilon$ expansions. This can be seen by comparing the results for the critical exponents in the regions of overlap of the expansions. Viewing $n$ and $d$ as continuous variables, there is therefore a continuous family of fixed points in a region of the $(n, d)$ plane for $2<d<4$ and sufficiently large $n$ \cite{March Russell}. This region is defined by $n>n_*(d)$, where $n_*(d)$ is the $d$-dependent value of $n$ at which the fixed point disappears. From the $4-\epsilon$ approach we know that the limiting behaviour as $d\rightarrow 4$ is $n_*(d)\rightarrow n_{*4} \simeq 183$, and from  the $2+\epsilon$ expansion we know that $n_*(d)\rightarrow 0$ as $d\rightarrow 2$. Using the sigma model we may also argue   that the slope of $n_*(d)$ is finite as $d\rightarrow 2$ (see endnote \cite{marginal operator in sigma model footnote}). The value of $n_*(3)$ is not known, but it is possible to show that $n_*(3)>1$ (Sec.~\ref{bound on n(3)}). We will assume that $n_*(d)$ increases monotonically with $d$.

We can go further by noting that in the $4-\epsilon$ expansion the way in which the nontrivial fixed point disappears at $n_*(d)$ is by annihilation with a tricritical point. On the basis of continuity we expect that the mechanism for the disappearance of the fixed point at $n_*(d)$ is the same for all $d$. This leads directly to the topology  in Fig.~\ref{rgflows3d}. It will be convenient to denote the value of $d$ where the merging of the critical and tricritical points happens, for a given $n$, by $d_*(n)$.

\subsubsection{RG flows close to merging line}

\noindent 
Since $n_{*4}\simeq 183$ is relatively large, the average slope of the line $d=d_*(n)$ is small.  It is therefore possible that there is a broad range of $n$--values where the line lies close to three dimensions, i.e. where $|d_*(n)-3|$ is small. So it is worth studying the RG flows in this regime. 

On the line, $\lambda$ is marginal. After rescaling and shifting $\lambda$ by a constant, its RG equation is:
\be
\f{\dd \lambda}{\dd \ln L} \simeq - \lambda^2.
\ee
 Moving slightly away from the line, the RG equation becomes, to lowest order in $d_*(n) - d$,
\be\label{RG equation}
\f{\dd \lambda}{\dd \ln L} \simeq  a(n) \big(  d_*(n) - d \big) - \lambda^2
\ee
with an unknown but universal positive constant $a(n)$. This equation encapsulates the fact that when ${d>d_*(n)}$ there is no fixed point, and when ${d<d_*(n)}$  both a critical and a tricritical point exist, at ${\lambda = \pm \sqrt{a(n) \left(  d_*(n) - d \right)}}$.

Fixing now on $d=3$, we define the universal quantity $\Delta(n) = a(n) (d_*(n)-3)$, which is zero at  $n=n_*(3)$ (where the critical point disappears in 3D) and small over the range of $n$ where the merging line lies close to $d=3$:
\be\label{RG equation 2}
\f{\dd \lambda}{\dd \ln L} \simeq   \Delta(n) - \lambda^2.
\ee

When $n< n_*(3)$, $\Delta(n)$ is negative and there is no fixed point in 3D. Instead the RG flows go off to large negative $\lambda$, suggesting a first order transition. However if we are close to the line, so $|\Delta(n)|$ is small, the RG flow becomes very slow. Integrating Eq.~\ref{RG equation 2} shows that this implies an exponentially large correlation length at this transition:
\be\label{correlation length}
\xi \sim \exp \lf \pi/\sqrt{|\Delta(n)|} \ri
\ee
with a nonuniversal prefactor. (A similar phenomenon occurs in the 2D $Q$-state Potts model for $Q\gtrsim 4$ \cite{cardy nauenberg scalapino}.)

For $n>n_*(3)$ there \emph{is} a conventional critical point at $\lambda = \sqrt{\Delta(n)}$. But if we are close to the line the leading irrelevant RG eigenvalue, $y_\text{irr}$, at this critical point is small, implying large corrections to scaling: from Eq.~\ref{RG equation},
\be
y_\text{irr} \simeq - 2 \sqrt{\Delta (n)}.
\ee

\subsubsection{Interpretation}

\noindent
At first sight the topology we have found for the RG flows suggests two possible explanations for the scaling violations. 

First, if we (speculatively) assume that $n_*(3) > 4$, but that the merging line lies close to $d=3$ over the range $2\leq n \leq 4$, then we obtain an anomalously weak first order transition for $n=2,3,4$. In this scenario there is pseudo-critical behaviour (with drifting exponents \cite{drifting exponents footnote}) up to the exponentially large lengthscale of Eq.~\ref{correlation length}, thanks to the  `nearby' fixed point at slightly smaller spatial dimension. The virtues of this scenario are that it appeals to universal features of the RG flow, so may explain why numerous different lattice models see very similar scaling violations, and that it can produce scaling violations over a range of $n$. 

(Note: if $|\Delta|$ is very small, there exists a range of sizes where $\lambda$ appears to be a conventional marginally irrelevant variable. However this range ends  at a size $L_*$ which is parametrically smaller than $\xi$ \cite{marginally irrelevant regime footnote} when $\xi$ is large.  The basic point is that the stages of the RG flow with $\lambda>0$ and with $\lambda<0$ take roughly equal amounts of RG time, but the latter corresponds to a vastly larger lengthscale because of the logarithmic relationship between length and RG time.)

Second, we might try to explain the scaling violations differently by postulating that in 3D the values $n=2,3,4$ lie just below the merging line ($n_*(3)<2$) so that $\Delta(n)$ is small and positive for $n=2,3,4$. This would give a true critical point with large (but conventional) scaling corrections due to a small irrelevant exponent $y_\text{irr}$. However, our numerical results strongly indicate that a small $y_\text{irr}$ is not sufficient, on its own, to explain what we see.

We emphasise that since our argument fixes the topology of the RG flows but not the numerical value of $n_*(d)$, the scenario above for a weak first order transition is speculative. In Sec.~\ref{interpretation of results} we will discuss other possible conjectures.

\subsubsection{Bound on $n_*(3)$}
\label{bound on n(3)}

\noindent
The value of $n_*(3)$ is not known, but we can argue that 
\be\label{nstar bound}
n_*(3) > 1.
\ee
This may look surprising at first glance, since the single-component Abelian Higgs model, $n=1$, certainly has a continuous phase transition in 3D. This transition is related by duality to that of the XY model \cite{dasgupta halperin}. Eq.~\ref{nstar bound} means that this `inverted XY' phase transition does not lie on the sheet we are considering: it is not analytically connected to the deconfined critical point at large $n$ \cite{Inverted XY footnote}.

Formally one can see this as follows. If the inverted XY transition did lie on the critical sheet of Fig.~\ref{rgflows3d}, we could describe it by setting $n=1$ in the $2+\epsilon$ expansion of the $\cp^{n-1}$ nonlinear sigma model. But this is evidently not the case. The inverted XY transition is a conventional thermodynamic phase transition with nontrivial signatures in the free energy. In contrast, the sigma model at  $n= 1$ is a replica-like theory, in which the number of degrees of freedom becomes zero and the free energy vanishes identically. The same reasoning implies that $n_*(3)$ is strictly greater than one. Otherwise the $n=1$ model would have a Higgs transition with an unphysical replica-like continuum description.

Instead of being connected to the critical points in Fig.~\ref{rgflows3d}, the inverted XY transition is the $n=1$ limit of a much simpler transition --- namely that  which (for $n>1$) separates the Higgs phase, where $\zb$ is condensed,  from a \emph{pair} condensed phase where the bilinear $\zb \zb^\dag$ is condensed but $\zb$ itself is  not. (This phase appears for appropriate couplings \cite{Kuklov et al, Herland}.) $\mathrm{SU}(n)$ symmetry is broken on \emph{both} sides of this transition, so it is not like the critical points in Fig.~\ref{rgflows3d}. One can check that the transition is in the inverted XY universality class for all $n$ (as the interactions between the critical sector and the Goldstone modes of the broken $\mathrm{SU}(n)$ symmetry are irrelevant), so the $n$ dependence of the critical behaviour is trivial \cite{more on n=1 footnote}.

\subsubsection{`Failure' of the $2+\epsilon$ expansion of the $\mathrm{O}(3)$ model}
\label{failure of 2+epsilon}

\noindent
The $\cp^1$ nonlinear sigma model is the $\mathrm{O}(3)$ nonlinear sigma model by another name, as the target space is the sphere ($\cp^1=S^2$).  Therefore the topology of the flows in Fig.~\ref{rgflows3d} confirms that the standard $2+\epsilon$ expansion of the $\mathrm{O}(3)$ sigma model does \emph{not} describe the Wilson Fisher critical point of the 3D $\mathrm{O}(3)$ model. Instead, setting $\epsilon=1$ in this expansion describes the $\mathrm{SU}(2)$ deconfined critical point, if it exists (or nothing at all if the critical point vanishes at a $d_*(2)$ below $3$). The applicability of the $2+\epsilon$ expansion to the abelian Higgs model was also argued for in Ref.~\cite{March Russell}.

Although this is contrary to what is often assumed, it should not be surprising in the light of knowledge about hedeghogs in the 3D $\mathrm{O}(3)$ model \cite{Kamal Murthy, Motrunich Vishwanath}. Suppressing these topological defects has been convincingly argued to change the critical behaviour \cite{Motrunich Vishwanath}. It is  natural that the $2+\epsilon$ expansion, which considers only spin waves,  describes the behaviour in the absence of hedgehogs \cite{higher gradient operators footnote}.

The conclusion is also supported by an RG approach to the $\mathrm{O}(M)$ model that employs a double expansion in $(d-2)$ and $(M-2)$ \cite{Cardy Hamber}. This shows that the standard $2+\epsilon$ expansion fails to capture the critical behaviour of the $\mathrm{O}(M)$ model when $M$ is smaller than a $d$-dependent critical value $M_c$. To first order in $d-2$, the relationship is $(M_c-2) \simeq \f{\pi^2}{4} (d-2)$, which gives $M_c\sim 4.5$ in 3D. While higher order corrections in $(d-2)$ could be significant, this result suggests $M_c>3$ \cite{Cardy Hamber}.

Ref.~\cite{Cardy Hamber} also notes the poor agreement between the $2+\epsilon$ exponents in 3D  and the exponents of the $\mathrm{O}(3)$ model in 3D. For example, $\nu$ is equal to $\nu=1/2$ at order $\epsilon^2$, or to $2/5$ at order $\epsilon^3$ \cite{brezin zinn-justin, hikami beta functions 1981}. In the light of the preceding, we should instead apply these exponents to the $\mathrm{SU}(2)$ deconfined transition. Indeed the $\epsilon$-expansion values $\nu = 1/2$ and $\nu = 2/5$ are remarkably close to our best estimates of $\nu$ (see Sec.~\ref{corr exp drift}).

\section{Interpretation and conclusions}
\label{interpretation of results}

\noindent
We have argued that the scaling violations at the deconfined critical point are too severe to be explained as corrections to scaling from a weakly or marginally irrelevant scaling variable, and have sharpened the possible alternatives. 

Some of our numerical results suggest that estimates of the exponents $\nu$, $\eta_\text{N\'eel}$ and $\eta_\text{VBS}$ may be better defined if we avoid  using finite-size scaling to obtain them (see in particular Secs.~\ref{correlation functions section},~\ref{corr exp drift}) --- although of course abandoning finite size scaling restricts us to lengthscales much smaller than the system size.  The most radical conjecture  would therefore be to attribute the scaling violations to a dangerously irrelevant variable  (DIV) which leaves critical behaviour intact but modifies finite-size scaling (see also Ref.~\cite{Kaul SU(3) SU(4)}). In the simplest picture, the role of this DIV would be to cut off fluctuations of some zero-mode(s) of the fields that are unbounded in the pure fixed-point theory. The main examples we know of this phenomenon are in free theories (such as $\phi^4$ theory above 4D \cite{brezin zinn justin fss} or the quantum Lifshitz theory \cite{quantum lifshitz}); another type of example is in theories that are dual to free theories \cite{duality footnote}. Further work is required to determine whether it is a plausible possibility in an interacting theory such as Eq.~\ref{NCCP1 lagrangian}.

The alternative possibility of an anomalously weak first order transition has been discussed in detail in Sec.~\ref{RG flow section}. We have seen that in principle there is a mechanism by which a very large correlation length can appear without the need for fine--tuning of the Hamiltonian, and that in this scenario there would be pseudo-critical behaviour over a large range of scales with (for example) drifting critical exponents. Such a possibility is hard to exclude numerically. We note however that we do not see the usual signs of an incipient first order transition, despite studying much larger scales than the early simulations used to argue for first-order behaviour in the J--Q model \cite{Jiang et al}. The first-order scenario would also leave the good scaling of, for example, the derivatives of the correlators (Sec.~\ref{correlation functions section}) a mystery.

Intriguing questions therefore remain for the future. The loop model is an ideal platform for further work on the deconfined transition, since it provides an intuitive geometrical picture and since isotropy in three dimensions is a convenient feature. It would also be interesting to perform simulations at other values of $n$ (which in the formulation just after Eq.~\ref{preliminary partition function} need not be an integer) in order to probe the scenario of Sec.~\ref{RG flow section}.

Various modifications to the loop model are possible. For example one may allow a third node configuration (see Fig.~\ref{nodes}) in which the two incoming links are joined and the two outgoing links joined, so that the loops are no longer consistently oriented. The symmetry is then broken from $\mathrm{SU}(n)$ to $\mathrm{SO}(n)$, and the relevant field theories are $\rp^{n-1}$ models \cite{cpn loops short, cpn loops long, loops with crossings, length distributions} which can show a $\mathbb{Z}_2$ spin liquid phase \cite{lammert, kaul SO(N)}. One may also study the effects of various anisotropies and symmetry-breaking perturbations. Finally, the loop model generalises to four dimensions, where it may be a useful tool to search for new types of critical behaviour arising in 3+1D quantum magnets.

\acknowledgements

We thank F. Alet, L. Balents, J. Cardy, R. Kaul, A. Ludwig, A. Sandvik, and T. Senthil for very useful discussions. This work was supported in part by EPSRC Grant No. EP/I032487/1 and by Spanish MINECO and FEDER (UE) grant no. FIS2012-38206 and MECD FPU grant no. AP2009-0668. AN acknowledges the support of a fellowship from the Gordon and Betty Moore Foundation under the EPiQS initiative (grant no. GBMF4303). MO thanks KITP for hospitality.

\appendix

\section{Methods}

The numerical procedure is as follows. 
An initial state is constructed by choosing at random one of the two possible configurations of a node with equal probabilities. 
Loops are then formed by following turning instructions at each node. We generate the fugacity $n$ from a sum on loop colours, assuming $n$ is integer. So  a colour is associated with each loop chosen with equal probability from $n$ alternatives.

Subsequent states are generated using three kinds of parallelized Monte Carlo moves to ensure that at equilibrium configurations are distributed according to the partition function Eq.~\ref{full partition function}. The first one updates the state of the nodes using a checkerboard-like algorithm. Each node, if its four links have the same colour, changes its state with probability $\min\left\{\exp\CC{-2J\s(i)\sum_{j_{nn}}\s(j)},1\right\}$. The updates of all the nodes in the lattice are done in three stages. In the L lattice there are two sublattices A and B. Each sublattice is tripartite and the three sublattices are simple cubic. At each time one sublattice of A and one sublattice of B are updated.
The second type of Monte Carlo move chooses a link at random and changes the colour of all the links of the associated loop to a different colour, chosen with uniform probability from the $n-1$ possibilities. This move is also parallelized by letting each thread choose a link and change the color of the loop if it is not already visited. The third type of move is to re-colour all loops in
the system, with the new colours selected independently
and at random for each loop. It is designed to ensure
that the colours of short loops equilibrate effciently.

{A number of these moves are combined to form a composite update which we term a Monte Carlo sweep. The ingredients in a single sweep are as follows. First we iterate 20 times a sequence in which moves of the first two types are intercalated and repeated for each of the three sublattices.  Then we apply the third type of move. Measurements are performed once every Monte Carlo sweep. The autocorrelation function of the energy is used to estimate a correlation time. The blocking and bootstrap methods \cite{newman book} are used to estimate errors.}

As an interpolation scheme we have used Ferrenberg's multiple histogram method \cite{Ferrenberg} to obtain a continuous set of values as a function of $J$ of the different quantities. This method is also employed whenever a derivative has to be calculated and a maximum or minimum has to be obtained. Errors related to this technique are calculated using the bootstrap method.

We consider system sizes of up to $3.9\times10^8$ links or lateral size $L=640$, with extensive results for $L\leq 512$. The minimum number of Monte Carlo sweeps used is $10^5$ for any $J$ and $L$, and increases with decreasing $L$.

\section{Calculation of hedgehog fugacity}
\label{hedgehog calculation}

Regarded as a  lattice magnet for classical $\cp^{n-1}$ spins, the partition function (\ref{lattice cpn-1 model}) has the peculiar feature that although it is both local and gauge invariant, it is not simply expressed in terms of the gauge invariant quantity $\vec N$ (for $n=2$) or $Q$ (for larger $n$). The consequence of this, as noted in Sec.~\ref{continuum description section}, is that the sigma model action arising from a derivative expansion may need to be supplemented by purely imaginary terms from nodes at which the phase of $\zb$ changes abruptly. In the presence of hedgehogs such nodes are inevitable, even far from the hedgehog core. As a result the hedgehog fugacity acquires a spatially varying phase.

For simplicity we take the hedgehogs to sit at the centre of a cube of $C_1$ or $C_2$ (Eq.~\ref{C1 and C2}), for example at the centre of the cube in Fig.~\ref{L lattice figure} (left). These locations form a bcc lattice. We take the origin at one such bcc site and the coordinate axes parallel to the links.

Focussing on the case $n=2$ (the generalisation to larger $n$ is immediate), let us first consider the representative configuration in which the hedgehog is centred at the origin and the N\'eel vector $\vec N$ (defined in Sec.~\ref{continuum description section}) points in the radial direction. In polar coordinates, $0\leq \theta\leq\pi$, $0\leq \varphi <2\pi$, this is
\be
\vec N = ( \sin \theta \cos \varphi, \sin\theta \sin \varphi, \cos\theta),
\ee
where the coordinates of a link are those of its midpoint. To write this in terms of $\vec z$ we must pick a gauge. It is convenient to choose one in which  the Boltzmann weight 
\be\label{appendix node weight}
e^{-S_\text{node}}= \f{1}{2} \left[ (\zb_o^\dag \zb_i)(\zb_{o'}^\dag \zb_{i'})+ (\zb_{o}^\dag \zb_{i'})(\zb_{o'}^\dag \zb_i) \right] 
\ee
is approximately equal to one for as many nodes as possible. For the links with positive and negative $z$ coordinate ($\theta<\pi/2$ and $\theta>\pi/2$) we take, respectively,
\ba \label{upper and lower hemispheres}
\zb & =  ( \cos\theta/2, e^{i\varphi} \sin \theta/2), &
\zb & = ( e^{-i\varphi} \cos\theta/2,  \sin \theta/2).
\end{align}
We see from Fig.~\ref{L lattice figure} that there are also links in the equatorial plane. For these we take
\be\label{equator}
\zb =  \f{1}{\sqrt 2}( e^{-i\varphi/2}, e^{i\varphi/2}).
\ee
Now consider $e^{-S_\text{node}}$. In fact it will suffice to consider only nodes far from the core, and to treat $\zb$ as constant except for the discontinuities in our gauge choice: other contributions to the action are either included in the spatially \emph{independent} amplitude of the hedgehog fugacity, or are already captured by the naive derivative expansion. 

With the above gauge choice, the nodes at which $\zb$ varies abruptly all lie in the equatorial plane, and have two of their links within this plane, one above it, and one below. Fig.~\ref{L lattice figure} shows four such nodes (all black). From Eqs.~\ref{appendix node weight}---\ref{equator} we find that most of these nodes have $e^{-S_\text{node}}\simeq 1$, despite the variation in the phase of $\zb$. However there is a string of nodes along the positive $x$ axis ($\varphi = 0$, $\theta=\pi/2$) each of which contributes a minus sign to the Boltzmann weight (i.e. $e^{-S_\text{node}}\simeq-1$ for these nodes). If we translate the core of the hedgehog by the vector $(2,0,0)$, while keeping the configuration far from the core fixed, we change number of nodes on this string by one. Therefore this translation changes the sign of the Boltzmann weight.

Let the phase term in the hedgehog fugacity be denoted
\be
\exp\lf i \, \eta(r) \ri,
\ee
where the spatial vector $r$ lies on a bcc site. (An `antihedgehog' of negative topological charge has phase factor $e^{-i \eta(r)}$, as we see from Eq.~\ref{appendix node weight} and the fact that complex conjugating $\vec z$ exchanges hedgehogs and antihedgehogs.) The phase $\eta(r)$ is defined only up to a constant: for example in a closed system there are equal numbers of hedgehogs and antihedgehogs, so the constant part of $\eta$ drops out. 

It may be seen from Fig.~\ref{L lattice figure} that the translational symmetry between bcc sites is not spoiled by the link orientations. Using this, we may argue that $\eta(r)$ is of the form $\eta(r) = k. r$ for some momentum $k$. By the above calculation, $e^{i \eta(r)} = - e^{i\eta(r + 2 \hat x)}$, where $\hat x= (1,0,0)$. By symmetry, we have similar results for translations in the $y$  and $z$ directions. This is enough to fix $k$ up to a sign:
\be
k = \pm  \f{\pi}{2} (1,1,1).
\ee
One of these signs applies to the hedgehog and one to the anti-hedgehog. We have not fixed which is which, but it does not matter.

With this $k$, the hedgehog fugacity takes four distinct values on the four sublattices of the bcc lattice, proportional to $\pm 1$ and $\pm i$. This is the result quoted in Sec.~\ref{continuum description section}.


\begin{thebibliography}{200}
\bibitem{anisotropy footnote} See e.g. Fig.~9 of Ref.~\cite{Kaul SU(3) SU(4)} which quantifies finite-size corrections to $z=1$ scaling in the $\mathrm{SU}(3)$ and $\mathrm{SU}(4)$ J--Q models.
\bibitem{N versus Ns footnote}  This measurement probes the fractal structure of loops on scales smaller than the system size. An alternative is to attempt a scaling collapse of the total length of system-spanning strands versus the number of such strands (more precisely of $N L^{3-d_f}$ against the spanning number $\spn$ defined in Sec.~\ref{spanning number section}; data not shown). This gives similar results: $\eta_\text{N\'eel}\sim 0.42$ if small $L$ values are included in the fit, falling to $\eta_\text{N\'eel} \sim 0.31$ when sizes less than $L=250$ are dropped.
\bibitem{O(5) footnote} This formulation \cite{senthil fisher competing orders} is a sigma model that groups the N\'eel and VBS order parameters into a single five-component vector. It includes a Wess-Zumino-Witten term, and anisotropy terms that break $\mathrm{O}(5)$ symmetry for rotations of this vector. If the transition was continuous and the anisotropies (with the exception of the operator driving the transition) happened to be irrelevant, $\mathrm{O}(5)$ symmetry would emerge at large scales. This might be more surprising in $2+1$D than in $1+1$D  \cite{1D chain footnote}, since in the absence of any symmetry-breaking terms the $\mathrm{O}(5)$ sigma model (with WZW term) already has one stable fixed point which is the ordered phase. If we postulate another stable critical fixed point, then we must also postulate an unstable multicritical point in between. However, this cannot be ruled out.
\bibitem{1D chain footnote} This is described by the $SU(2)_1$ Wess-Zumino-Witten model, which has an emergent $\mathrm{O}(4)$ symmetry that rotates the dimerisation and the three components of the N\'eel vector into each other  \cite{tsvelik book}.
\bibitem{spanning number footnote} With open BCs, $\<\mathcal{N}_s\>$ is identical  to the stiffness defined in the usual way by the response of the free energy to an infinitesimal twist in the BCs for the N\'eel vector \cite{cpn loops long, loops with crossings}. Here we use periodic BCs and define $\mathcal{N}_s$ by slicing open the system. This means that $\<\mathcal{N}_s\>$ is no longer simply related to the free energy, but does not change the qualitative scaling behaviour.
\bibitem{relative phase footnote} E.g. in a finite system with a single hedgehog, the absolute phase depends on the boundary conditions for $\vec z$. With periodic BCs we have an equal number of hedgehogs and antihedgehogs. An antihedgehog (of opposite topological charge) has the opposite phase.
\bibitem{unimportance of fourfold anisotropy footnote} It is also now clear the scaling violations are not due to fourfold anisotropy in the VBS order parameter (see Refs.~\cite{pujari damle alet honeycomb, block melko kaul, lou sandvik kawashima} and Sec.~\ref{emergent symmetry section}), so the explanation for them should be sought in the framework of the $\nccp^1$ model.
\bibitem{marginal operator in sigma model footnote} The merging with the tricritical point occurs when a scaling variable becomes marginal. In the NLSM,  the only candidates that preserve the symmetry are operators with more than two derivatives. Those with four derivatives have RG eigenvalues of the form $y = f(t^*)-d$, where $t_*$ is the sigma model coupling at the critical point \cite{Lerner Wegner}. Therefore $t_*$ is $O(1)$ at the point where $y=0$. For small $\epsilon$, the coupling $t_*$ is of order $\epsilon/n$. This indicates that $n_*(2+\epsilon) \propto \epsilon$ as stated in the text. Since $t_*$ is $O(1)$ in the regime of interest, lowest-order calculations do not fix the coefficient of proportionality.
\bibitem{drifting exponents footnote} One may also consider the RG equation for the leading thermal scaling variable which drives the transition. Very close to the merging line we can approximate this as $\dd \mu/\dd \ln L = (y+ b \lambda) \mu$ (similarly to e.g. \cite{cardy nauenberg scalapino, cpn loops long}) which gives one way to define an effective $\nu=(y+ b \lambda)^{-1}$ that drifts with $\lambda$.
\bibitem{marginally irrelevant regime footnote} $\lambda$ will appear to be a conventional marginally irrelevant variable only when $\lambda \gtrsim |\Delta|^{1/2}$, so that the second term in Eq.~\ref{RG equation 2} dominates. For example, defining $L_*$ as the scale where $\lambda^2 = |\Delta|$ gives $L_* \sim \xi^{1/4}$.
\bibitem{Inverted XY footnote} The assumption is often made in the literature that the scaling dimensions of monopole insertion operators  interpolate smoothly, as a function of $n$, between the value at the inverted XY transition and in the large $n$ regime. In the light of the above, this assumption is not justified.
\bibitem{higher gradient operators footnote} The reasons given here for the failure of the $2+\epsilon$ expansion of the $\mathrm{O}(3)$ model should not be confused with earlier speculations that for \emph{any} fixed $\epsilon$, operators with a very large number of derivatives may become relevant and destabilize the fixed point. Ref~\cite{brezin hikami irrelevance} has argued that such speculations are not well-founded.
\end{thebibliography}
\end{document}